\documentclass[format=acmsmall, review=false, screen=true]{acmart}

\usepackage{booktabs} % For formal tables

\usepackage{makecell}
\usepackage{array}
\usepackage{varwidth}

\usepackage[ruled]{algorithm2e} % For algorithms

\SetAlFnt{\small}
\SetAlCapFnt{\small}
\SetAlCapNameFnt{\small}
\SetAlCapHSkip{0pt}
\IncMargin{-\parindent}

\usepackage{cleveref}
\usepackage[utf8]{inputenc}
\crefname{section}{§}{§§}
\Crefname{section}{§}{§§}
\usepackage[10pt]{moresize}
\usepackage[font={sf}]{caption}
\usepackage[font={sf}]{subcaption}
\usepackage{amsmath,amssymb,amsfonts}
\usepackage{mathtools,mathrsfs}
\usepackage{multirow}
\usepackage{rotating}
\usepackage{enumitem}
\newcommand{\macb}[1]{\textbf{\textsf{#1}}}
 \usepackage[hang,flushmargin]{footmisc} 
 
 \newcommand\blfootnote[1]{%
   \begingroup
   \renewcommand\thefootnote{}\footnote{#1}%
   \addtocounter{footnote}{-1}%
   \endgroup
 }
\usepackage{scalerel,stackengine}
\stackMath
\newcommand\rwh[1]{%
\savestack{\tmpbox}{\stretchto{%
  \scaleto{%
      \scalerel*[\widthof{\ensuremath{#1}}]{\kern-.6pt\bigwedge\kern-.6pt}%
          {\rule[-\textheight/2]{1ex}{\textheight}}%WIDTH-LIMITED BIG WEDGE
            }{\textheight}% 
}{0.5ex}}%
\stackon[1pt]{#1}{\tmpbox}%
}

\usepackage{graphicx}

\usepackage{fontawesome}
\usepackage{pifont}

\usepackage{tikz}
\usetikzlibrary{tikzmark}

\usepackage[per-mode=symbol]{siunitx}
\DeclareSIUnit \bit {bit}
\DeclareSIUnit \byte {B}
\DeclareSIUnit \Byte {Byte}
\DeclareSIUnit \cycle {cycle}
\DeclareSIUnit \cycles {cycles}
\DeclareSIUnit \hz {Hz}
\DeclareSIUnit \op {Op}
\DeclareSIUnit \operand {operand}
\DeclareSIUnit \operands {operands}
\DeclareSIUnit \transfer {T}

\usepackage{xpatch}
\expandafter\xpatchcmd
   \csname pgfk@/tikz/every picture/.@cmd\endcsname
            {\thepage}{\arabic{page}}{}{}

            \tikzstyle{comment} = [draw, fill=blue!70, text=white, text width=3cm, minimum height=1cm, rounded corners, align=left, font=\scriptsize]
            \tikzstyle{background_alg} = [draw, fill=blue!20, opacity=0.4, inner sep=4pt, rounded corners=2pt]

            \usetikzlibrary{shapes}
            \usetikzlibrary{plotmarks}
            \usetikzlibrary{calc,fit}

\acmDOI{0000001.0000001}

\begin{document}
% Title portion. Note the short title for running heads 

\newcolumntype{M}{>{\begin{varwidth}{8cm}}l<{\end{varwidth}}} %M is for Maximal column

%\title[Survey and Taxonomy of Graph Compression]{Survey and Taxonomy of Lossless and Lossy Graph Compression}
\title{Graph Processing on FPGAs: Taxonomy, Survey, Challenges}
%\title[Survey and Taxonomy of Lossless Graph Compression]{Survey and Taxonomy of Lossless Graph Compression}
\subtitle{Towards Understanding of Modern Graph Processing, Storage, and Analytics}

\author{Maciej Besta*, Dimitri Stanojevic*}\blfootnote{*Both authors contributed equally to the work}
\affiliation{
  \institution{Department of Computer Science, ETH Zurich}
  \country{Switzerland}
}
 
% \author{Maciej Besta}
% \affiliation{
%   \institution{Department of Computer Science, ETH Zurich}
%   \country{Switzerland}
% }
% 
\author{Johannes de Fine Licht, Tal Ben-Nun}
\affiliation{
  \institution{Department of Computer Science, ETH Zurich}
  \country{Switzerland}
}

\author{Torsten Hoefler}
\affiliation{
  \institution{Department of Computer Science, ETH Zurich}
  \country{Switzerland}
}

\renewcommand{\shortauthors}{M. Besta, D. Stanojevic, J. de Fine Licht, T. Ben-Nun, T. Hoefler}

\begin{abstract}
Graph processing has become an important part of various areas, such as machine
learning, computational sciences, medical applications, social network
analysis, and many others.  Various graphs, for example web or social networks, may
contain up to trillions of edges. The sheer size of such datasets, combined
with the irregular nature of graph processing, poses unique challenges for the
runtime and the consumed power. 
Field Programmable Gate Arrays (FPGAs) can be an energy-efficient solution to
deliver specialized hardware for graph processing. This is reflected by the
recent interest in developing various graph algorithms and graph processing
frameworks on FPGAs.
To facilitate understanding of this emerging domain, we present the first
survey and taxonomy on graph computations on FPGAs. Our survey describes and
categorizes existing schemes and explains key ideas.
% 
% Moreover, considering the fact that overcoming the memory bottleneck is a
% large challenge in FPGA-related works, we also list the sizes of largest
% processed graphs in considered works.
%
Finally, we discuss research and engineering challenges to outline the future
of graph computations on FPGAs. 
\end{abstract}



\maketitle

% \renewcommand{\shortauthors}{Maciej Besta and Torsten Hoefler}

\section{Introduction}

Graph processing underlies many computational problems in social network
analysis, machine learning, computational science, and
others~\cite{DBLP:journals/ppl/LumsdaineGHB07, jiang2011short}. Designing
efficient graph algorithms is challenging due to several properties of graph
computations such as irregular communication patterns or little locality. These
properties, combined with the sheer size of graph datasets (up to trillions of
edges~\cite{ching2015one}), make graph processing and graph algorithms consume
large amounts of energy.

Most graph algorithms are communication-heavy rather than compute-heavy:
more time is spent on accessing and copying data than on the actual
computation. For example, in a Breadth-First Search (BFS)
traversal~\cite{Cormen:2001:IA:580470}, a fundamental graph algorithm, one
accesses the neighbors of each vertex. In many graphs, for example various
social networks, most of these neighborhoods are small (i.e., contain up to
tens of vertices), while some are large (i.e., may even contain more than
half of all the vertices in a graph). General purpose CPUs are not ideal for
such data accesses: They have fixed memory access granularity based on cache line sizes,
do not offer flexible high-degree parallelism, and their caches do not work
  effectively for irregular graph processing that have little or no temporal
  and spatial locality. GPUs, on the other hand, offer massive parallelism, but exhibit significantly reduced performance when the internal cores do not execute the same instruction (i.e., warp divergence), which is common in graphs with varying degrees. 

Field Programmable Gate Arrays (FPGAs) are integrated circuits that can be
reprogrammed using hardware description languages. This allows for rapid
prototyping of application-specific hardware. An FPGA consists of an array of
logic blocks that can be arbitrarily rewired and configured to perform
different logical operations. FPGAs usually use low clock frequencies of
$\approx$100--200MHz, but they enable building custom hardware optimized for a given
algorithm. Data can directly be streamed to the FPGA without the need to decode
instructions, as done by the CPU. This data can then be processed in pipelines
or by a network of processing units that is implemented on the FPGA, expressing parallelism at a massive scale.
% 
% This also allows to tailor
% the memory access granularity to the needs of a specific algorithm and thus
% avoid reading whole cache lines that may not be needed. \johannes{Unfortunately,
% because of DRAM technology, we actually might as well read cache lines on FPGAs
% also, so this problem doesn't really go away as long as we use DDR.} 
%
% The power efficiency, reconfigurability and low cost of FPGAs has also made
% them appealing for use as accelerators in large data centers
% \cite{putnam2014reconfigurable}.
%
Another major advantage of FPGAs is the large cumulated bandwidth of their
on-chip memory. Memory units on the FPGA, such as block RAM (BRAM), can be used
to store reusable data to exploit temporal locality, avoiding expensive
interactions with main memory. On a Xilinx Alveo U250 FPGA, 2566 memory blocks
with $\SI{72}{\bit}$ ports yield an on-chip bandwidth of
$\SI{7}{\tera\byte\per\second}$ at $\SI{300}{\mega\hertz}$, compared to
$\SI{1}{\tera\byte\per\second}$ for full 256-bit AVX throughput at maximum turbo
clock on a 12-core Intel Xeon Processor E5-4640 v4 CPU. In practice, the
advantage of FPGAs can be much higher, as buffering strategies are programmed
explicitly, as opposed to the fixed replacement scheme on a CPU.
%
%\johannes{You
%can decide if you want to include this. This is essentially peak performance of
%the CPU, converted into bandwidth from L1 cache. In practice this is much harder
%to achieve on a CPU than on FPGA.} 

Developing an application-specific FPGA accelerator usually requires more effort
than implementing the same algorithm on the CPU or GPU.  There are also many
other challenges. For example, modern FPGAs contain in the order of tens
of $\si{\mega\byte}$ of BRAM memory, which is not large enough to hold entire
graph data sets used in today's computations. 
%
%\johannes{You can make the same
%argument for CPU cache and GPU shared memory} 
%
Therefore, BRAM must be used as
efficiently as possible, for example by better optimizing memory access
patterns.
% 
% or by using more cache-friendly data structures. \johannes{Cache is
% maybe not the right word here, the point is that we don't have to use caches} 
%
Thus, a significant amount of research has been put into developing both
specific graph algorithms on FPGAs and graph processing frameworks that
allow to implement various graph algorithms easier, without having to develop
everything from scratch~\cite{khoram2018accelerating, yao2018efficient,
zhang2018degree, yang2018efficient, zhou2018framework, lee2017extrav,
ma2017fpga, zhou2017accelerating, zhang:graph_FPGA, dai:foregraph, lei2016fpga,
ozdal2016energy, zhou2016high, engelhardt2016gravf, oguntebi:GraphOps,
dai:fpgp}.

This paper provides the first taxonomy and survey that attempts to cover all
the associated areas of graph processing on FPGAs. Our goal is to (1)
\emph{exhaustively} describe related work, (2) illustrate and explain the
\emph{key ideas}, and (3) \emph{systematically} categorize existing algorithms,
schemes, techniques, methodologies, and concepts.
%
%\macb{What Is The Scope of This Survey?}
%
We focus on all works researching graph computations on FPGAs, 
both general frameworks as well as implementations of specific graph 
algorithms. 

\macb{What Is the Scope of Existing Surveys?}
To the best of our knowledge, as of yet there is no other survey on FPGAs for
graph processing. Only Horawalavithana~\cite{horawalavithana:graph} briefly
reviews several hardware accelerators and frameworks for graph computing and
discusses problems and design choices. However, the paper only partially
focuses on FPGAs and covers only a few selected works. 

\section{Background}

We first present concepts used in all the sections and summarize the key 
symbols in Table~\ref{tab:symbols}.

\begin{table}[h!]
%\vspace{-1em}
\centering
%\small
\footnotesize
%\scriptsize
%\ssmall
\sf
\begin{tabular}{ll@{}}
\toprule
                    $G, \mathbf{A}$ & A graph $G=(V,E)$ and its adjacency matrix; $V$ and $E$ are sets of vertices and edges.\\
                    $n,m$&Numbers of vertices and edges in $G$; $|V| = n, |E| = m$.\\
                    $d, \overline{d}, D$&Average degree, maximum degree, and the diameter of $G$, respectively.\\
                    $d_v, N_v$ & The degree and the sequence of neighbors of a vertex $v$.\\
                    $B_{DRAM}$ & The bandwidth between the FPGA and DRAM. \\
                    $B_{BRAM}$ & The bandwidth of a BRAM module.\\

%                    $\mathcal{O}, \mathcal{A}, \mathcal{A}_v$ & \makecell[l]{Data structures for, respectively, the pointers to the adjacency data of each vertex,\\the adjacency data of a given graph $G$, and the adjacency data of vertex $v$.} \\
%                    $N_{in,v}, N_{out,v}$ & The in-neighbors and out-neighbors of a vertex $v$.\\
%                    $N_{i,v}$ & The $i$th neighbor of $v$ (in the order of increasing labels).\\
%                    \midrule
% \multirow{3}{*}{\begin{turn}{90}\shortstack{Various}\end{turn}}
%                      $T,W$ &The number of threads; work complexity of a given scheme.\\
%                      $C,n_c$ &Chunk height; the number of chunks in SlimSell or Sell-$C$-$\sigma$.\\
%                      $\sigma$ &Sorting scope in SlimSell and Sell-$C$-$\sigma$ ($\sigma \in [1,n]$).\\
\bottomrule
\end{tabular}
\caption{The most important symbols used in the paper.}
%\vspace{-2em}
\label{tab:symbols}
\vspace{-1.5em}
\end{table}

%\vspace{-0.5em}
\subsection{Graphs}

We model an undirected graph~$G$ as a tuple $(V,E)$; 
$V$ is a set of vertices and $E \subseteq V \times V$ is a set of edges;
$|V|=n$ and $|E|=m$. If $G$ is directed, we use the name \emph{arc} to refer to
an edge with a specified direction.
An edge between vertices $v$ and $w$ is denoted as $(v,w)$.
We consider both labeled and unlabeled graphs. If a graph is labeled, $V = \{1,
..., n\}$, unless stated otherwise.  We use the name ``label'' or ``ID''
interchangeably.
%
% $A \subseteq V \times V $ denotes the set of all arcs. 
% 
$N_v$ and $d_v$ are the neighbors and the degree of a vertex $v$. 
% 
% The $i$th neighbor of $v$ (in the order of increasing labels) is denoted as
% $N_{i,v}$; $N_{0,v} \equiv v$.
%
% The (non-negative) weight of an edge $(v,w)$ is
% {\small$\mathcal{W}_{(v,w)}$}. 
% 
% We denote the maximum degrees for a given $G$ as $\overline{d}$, $\overline{d}_{in}$
% (in-degree), and $\overline{d}_{out}$ (out-degree).
%
% The average degree is denoted with a bar ($\overline{d}$). 
%
$G$'s diameter is $D$.
A subgraph of $G$ is a graph $G' = (V', E')$ such that $V' \subseteq V$ and $E'
\subseteq E$. In an \emph{induced} subgraph, $E'$ contains only edges
$(v,w)$ such that both $v$ and $w$ are in $V'$.
A path in $G$ is a sequence of edges in $G$ between two vertices
$v$ and $w$: $(v,v_1)$, $(v_1, v_2)$, ..., $(v_{n-1}, v_n)$, $(v_n, w)$.

% Each vertex $v$ is assigned a label $\lambda(v) \in \mathbb{N}$. Labels are
% unique ($v \neq u \Rightarrow \lambda(v) \neq \lambda(u)$), not necessarily
% contiguous, and can be freely reassigned. We identify vertices with their
% labels ($v \equiv \lambda(v)$) when it does not cause confusion.

% We focus on three fundamental queries~\cite{succinct_bound} on the graph
% structure: \textsf{deg(v)} and \textsf{neigh(v)} return the degree and the
% neighbors of a given vertex $v$; \textsf{adj(v,w)} returns \textsf{true} if two
% vertices $v$ and $w$ are neighbors, and \textsf{false} otherwise.

\subsection{Graph Processing Abstractions}
\label{sec:back-abstractions}

Graph algorithms such as BFS can be viewed in either the \textbf{traditional
combinatorial abstraction} or in the \textbf{algebraic
abstraction}~\cite{besta2017slimsell, kepner2016mathematical}. In the former,
graph algorithms are expressed with data structures such as queues or bags and
operations on such structures such as inserting a vertex into a
bag~\cite{leiserson2010work}. In the latter, graph algorithms are expressed
with basic linear algebra structures and operations such as a series of
matrix-vector (MV) or matrix-matrix (MM) products over various
semirings~\cite{kepner2011graph}.
%
% These matrix- products consist of a sparse matrix
% and a dense vector (SpMV) or a sparse matrix and a sparse vector (SpMSpV).
%
Both abstractions have advantages and disadvantages in the context of
graph processing. For example, BFS based on MV uses no explicit locking~\cite{schmid2016high} or
atomics~\cite{schweizer2015evaluating} and has a succinct description. Yet, it may need
more work than the traditional BFS~\cite{yang2015fast}.

\subsection{Graph Representations and Data Structures}

We discuss various graph-related structures used in FPGA works.

\subsubsection{Adjacency Matrix, Adjacency List, Compressed-Sparse Row}

$G$ can be represented as an \textbf{adjacency matrix} (AM) or
\textbf{adjacency lists} (AL). AL uses $\mathcal{O}(n \log n)$ bits and AM
uses $\mathcal{O}\left(n^2\right)$ bits.
%
% {AL} needs $O\left(\overline{d}\right)$ time to check if two vertices are connected
% while obtaining $N_v$ or $d_v$ takes $O(1)$ time. For {AM}, it takes $O(1)$ to
%   verify if two vertices are connected and $O\left(\overline{d}\right)$ to obtain
%   $N_v$ and $d_v$.
%
% \textsf{AL} needs $O(\overline{d})$ time to answer \textsf{adj(v,$\cdot$)} while
% \textsf{neigh(v)} and \textsf{deg(v)} take $O(1)$ time. Fetching the offset of
% the adjacency data of vertex $v$ takes $O(1)$ time and is referred to as
% \textsf{off($v$)}.
% 
% \textsf{AM} is not commonly used because it uses $O(n^2)$ bits of space and, in
% its uncompressed form, offers storage advantages over \textsf{AL} only for very
% dense graphs where $2m \log n + n \log m > n^2$.  For \textsf{AM}, it takes
% $O(1)$, $O(\overline{d})$, $O(\overline{d})$, and $O(1)$ time to answer the \textsf{adj},
% \textsf{neigh}, \textsf{deg}, and \textsf{off} queries, respectively.

When using AL, a graph is stored using a contiguous array with the adjacency
data and a structure with offsets to the neighbors of each vertex. 
When using AM, the graph can be stored using the well-known
\textbf{Compressed-Sparse Row} (CSR) format~\cite{saad1990sparskit}. In CSR,
the corresponding AM is represented with three arrays: $val$, $col$, and $row$.
$val$ contains all matrix non-zeros (that correspond to $G$'s edges) in the row
major order. $col$ contains the column index for each corresponding value in
$val$; it has the same size ($O(m)$). Finally, $row$ contains starting indices
in $val$ (and $col$) of the beginning of each row in the AM (size $O(n)$).
%
% To indicate on which row edges are located, adjacent entires in $row$ index
% into $val$ and $col$ as begin and end of all the vertices in a given row.

\subsubsection{Sparse and Dense Data Structures}
\label{sec:back-sparse-dense}

Data structures used in graph processing, such as bitvectors, can be either
\textbf{sparse} or \textbf{dense}. In the former, one only stores
\emph{non-zero elements} from a given structure, together with any indices
necessary to provide locations of these elements.  In the latter, \emph{zeros} are also stored.
In the case of MV, the adjacency matrix is usually sparse (e.g., when using
CSR) while the vector can be sparse or dense, respectively resulting in
\textbf{sparse-sparse (SpMSpV)} and \textbf{sparse-dense (SpMV)} MV products. The
latter entail more work but offer more potential for vectorization; the former
is work-efficient but has more irregular memory accesses.

\begin{table*}[t]
\vspace{-1em}
\centering
 \setlength{\tabcolsep}{1pt}
 \renewcommand{\arraystretch}{1}
%\footnotesize
\scriptsize
%\ssmall
%\sf
\begin{tabular}{llllll@{}}
\toprule
\multicolumn{2}{c}{\makecell[c]{\textbf{Graph problem + time complexity of the best}\\\textbf{(or established) sequential algorithm(s)}}} & \makecell[c]{\textbf{Associated}\\\textbf{graph/vertex}\\\textbf{property}} & \multicolumn{2}{l}{\makecell[l]{\textbf{Associated example algorithm and its time/work}\\\textbf{complexity (in the PRAM CRCW model~\cite{beame1989optimal})}}} & \makecell[c]{\textbf{Selected}\\\textbf{application}} \\
\midrule
%
% \multirow{6}{*}{\begin{turn}{90}\textbf{Distance}\end{turn}}
%
\makecell[l]{Single-Source Shortest Path\\(SSSP) (unweighted)~\cite{Cormen:2001:IA:580470}} & $O\left(m+n\right)$~\cite{Cormen:2001:IA:580470} & \makecell[l]{Length of a\\shortest path} & BFS~\cite{besta2017push} & $O\left(D \overline{d} + D \log m\right)$, $O\left(m\right)$  & \makecell[l]{Bipartite\\testing} \\
SSSP (weighted)~\cite{Cormen:2001:IA:580470} & \makecell[l]{$O\left(m + n \log n\right)$~\cite{fredman1987fibonacci},\\$O(m)$~\cite{thorup1999undirected}} & \makecell[l]{Length of a\\shortest path} & \makecell[l]{$\Delta$--Stepping~\cite{meyer2003delta},\\Bellman-Ford~\cite{bellman1958routing}} & \makecell[l]{$O\left(\overline{d} \cdot \overline{W_{{P}}} \cdot \log n + \log^2 n\right)^{\text{\textdagger}}$,\\$O\left(n + m + \overline{d} \cdot \overline{W_{{P}}} \cdot \log n\right)^{\text{\textdagger}}$}  & \makecell[l]{Robotics,\\VLSI design} \\
\makecell[l]{All-Pairs Shortest Path\\(APSP) (unweighted)~\cite{Cormen:2001:IA:580470}} & \makecell[l]{$O\left(mn + n^2\right)$~\cite{Cormen:2001:IA:580470},\\$O(n^3)$~\cite{floyd1962algorithm}} & \makecell[l]{Length of a\\shortest path} & \makecell[l]{BFS~\cite{besta2017push}} & \makecell[l]{$O\left(D \overline{d} + D \log m\right)$, $O\left(nm\right)$} & \makecell[l]{Urban\\planning} \\
\makecell[l]{All-Pairs Shortest Path\\(APSP) (weighted)~\cite{Cormen:2001:IA:580470}} & \makecell[l]{$O\left(mn + n^2 \log n\right)$~\cite{fredman1987fibonacci}} & \makecell[l]{Length of a\\shortest path} & \makecell[l]{Han et al.~\cite{Han1997}} &\makecell[l]{$O\left(n^2\right)$, $O\left(n^3\right)$} & \makecell[l]{Traffic\\routing} \\
\midrule
%
%
%\vspace{0.25em} \\
%
% & \makecell[l]{Diameter Estimation (DE),\\Radius Estimation (RE)~\cite{shun2013ligra}} & $O\left(n^2 + nm\right)$~\cite{shun2013ligra} & Diameter, radius & Multiple ($K$) BFS runs~\cite{shun2013ligra} & $O\left(K D\overline{d} + K D\log m\right)$, $O\left(Km\right)$  & Transportation \\
%
% \midrule
%
% \multirow{5}{*}{\begin{turn}{90}\textbf{Connectivity}\end{turn}}
%
\makecell[l]{[Weakly, Strongly]\\Connected Components~\cite{Cormen:2001:IA:580470},\\Reachability} & $O\left(m+n\right)$~\cite{Cormen:2001:IA:580470} & \makecell[l]{\#Connected\\components,\\Reachability} & Shiloach-Vishkin~\cite{shiloach1980log} & $O\left(\log n\right)$, $O\left(m \log n\right)$ & \makecell[l]{Verifying\\connectivity} \\
\midrule
%
% & $k$-Core Decomposition (CD)~\cite{khaouid2015k} & $O\left(m\right)$~\cite{batagelj2003m} & $k$-core-number & Ligra kernel~\cite{shun2013ligra, dhulipala2017julienne} & $O\left(\rho \log n\right)^{\text{\textdagger}}$, $O\left(m+n\right)^{\text{\textdagger}}$  & Analysis of proteins
%
% \vspace{0.25em} \\
%
Triangle Counting (TC)~\cite{shun2015multicore} & \makecell[l]{$O\left(m \overline{d}\right)$,\\$O\left(m^{3/2}\right)$~\cite{schank2007algorithmic}} & \#Triangles & GAPBS kernel~\cite{beamer2015gap} & $O\left(\overline{d}^2\right)$, $O\left(m \overline{d}\right)$  & \makecell[l]{Cluster\\analysis} \\
\midrule
%
%
% \vspace{0.25em} \\
%
% & \makecell[l]{Low-Diameter Decomposition (LDD)~\cite{miller2015improved}} & $O\left(m\right)$~\cite{miller2015improved} & -- & Miller et al.~\cite{miller2015improved} & $O\left(\log n \log^* n\right)$, $O\left(m\right)$   & Distance oracles \\
% & Low-Diameter Decomposition (LDD)~\cite{miller2015improved} (weighted) & $O\left(m\right)$ & Miller et al.~\cite{miller2015improved} & $O\left(\log U \log n \log^* n\right)$ & $O\left(m\right)$    & Distance oracles \\
%
% \midrule
%
% \multirow{6}{*}{\begin{turn}{90}\textbf{Optimization}\end{turn}}
%
\makecell[l]{Minimum Spanning\\Tree (MST)~\cite{Cormen:2001:IA:580470}} & \makecell[l]{$O\left(m \log n\right)$~\cite{Cormen:2001:IA:580470},\\$O(m \alpha(m,n))$~\cite{chazelle2000minimum}} & MST weight & Boruvka~\cite{boruuvka1926jistem} & $O\left(\log n\right)$, $O\left(m \log n\right)$  & \makecell[l]{Design of\\networks} \\
\midrule
\makecell[l]{Maximum Weighted\\Matching (MWM)~\cite{papadimitriou1998combinatorial}} & $O\left(m n^2\right)$ & MWM weight & Blossom Algorithm~\cite{edmonds1965paths} & --- & \makecell[l]{Comp.\\chemistry} \\
%
%\midrule
%
%\makecell[l]{Minimum Vertex\\Cover (VCover)~\cite{Cormen:2001:IA:580470}} & \talbn{Approximations?} & Vertex & \cite{ghaffari18improved}? &  & \makecell[l]{Race cond.\\detection} \\
%
\midrule
%
% & Maximum Flow (MF)~\cite{Cormen:2001:IA:580470} & $O\left(n^2 m\right)$~\cite{goldberg1988new}, $O\left(n m\right)$~\cite{orlin2013max, king1994faster} & MF value & Preflow--push~\cite{goldberg1988new} & $O\left(n^2 \log n\right)$ (EREW PRAM), unspecified  & Airline scheduling \\
% & Balanced Minimum Cut (MC)~\cite{Cormen:2001:IA:580470} & NP-Complete & MC size & METIS kernel~\cite{karypis1995metis} & unspecified & Network robustness \\
% & Maximum Independent Set (MIS)~\cite{papadimitriou1998combinatorial} & NP-Hard & MIS size & ECL-MIS~\cite{kulkarni2007optimistic} & unspecified, unspecified & Scheduling \\
% & Maximum Independent Set (MIS)~\cite{papadimitriou1998combinatorial} & NP-Hard & MIS size & Luby~\cite{luby1986simple} & $O\left(\log n\right)^{\text{\textdagger}}$, $O\left(m\log n\right)^{\text{\textdagger}}$ & Scheduling \\
%
%\vspace{0.25em} \\
%
% & Maximum Weighted Bipartite Matching (MWBM)~\cite{papadimitriou1998combinatorial} & $O\left(n^2 \log n + nm\right)$ & Blossom Algorithm~\cite{edmonds1965paths} & unspecified & unspecified & Transportation
%
%\vspace{0.25em} \\
%
%  & Minimum Graph Coloring (MGC)~\cite{papadimitriou1998combinatorial} & NP-Hard & Chromatic number & Jones and Plassmann~\cite{jones1993parallel} & $O\left(\log n / \log \log n\right)^{\text{\textdagger}}$, unspecified & Scheduling \\
%
% \midrule
%
% \multirow{8}{*}{\begin{turn}{90}\textbf{Centralities}\end{turn}}
%
\makecell[l]{Betweenness Centrality\\(BC)~\cite{brandes2001faster} (unweighted)} & $O\left(n m\right)$~\cite{brandes2001faster} & Betweenness & Parallel Brandes~\cite{brandes2001faster, beamer2015gap} & \makecell[l]{$O\left(n D \overline{d} + n D \log m\right)$,\\$O\left(nm\right)$}  & \makecell[l]{Network\\analysis} \\
%
% \vspace{0.25em} \\
%
BC~\cite{brandes2001faster} (weighted) & $O\left(n m + n^2 \log n\right)$~\cite{brandes2001faster} & Betweenness & Parallel Brandes~\cite{brandes2001faster, beamer2015gap} & ---  & \makecell[l]{Network\\analysis} \\
\midrule
%
%
% \vspace{0.25em} \\
%
% & Triangle Counting per Vertex (TCV)~\cite{shun2015multicore} & $O\left(n \overline{d}^2\right), O\left(m^{3/2}\right)$ & \#Triangles & GAPBS kernel~\cite{beamer2015gap} & $O\left(\overline{d}^2\right)$, $O\left(m \overline{d}\right)$  & Cluster analysis
%
% \vspace{0.25em} \\
%
Degree Centrality (DC)~\cite{Cormen:2001:IA:580470} & $O\left(m+n\right)$~\cite{Cormen:2001:IA:580470} & Degree & Simple listing~\cite{Cormen:2001:IA:580470} & $O\left(1\right)$, $O\left(m+n\right)$  & \makecell[l]{Ranking\\vertices} \\
\midrule
%
%
% \vspace{0.25em} \\
%
PageRank (PR)~\cite{page1999pagerank} & $O\left(I m\right)$~\cite{beamer2015gap} & Rank & GAPBS kernel~\cite{beamer2015gap} & $O\left(I \overline{d}\right)$, $O\left(I m\right)$  & \makecell[l]{Ranking\\websites} \\
\bottomrule
\end{tabular}
%
%\vspace{-1.5em}
%
\caption{\textbf{Overview of fundamental graph
problems and algorithms considered in FPGA works.} $^{\text{\textdagger}}$Bounds in expectation or with
high probability.
$\alpha(n,m)$ is the inverse Ackermann function.
$\overline{W_{{P}}}$ is the maximum shortest path weight between any two
vertices. $I$ is the number
of iterations in PageRank.
}
% 
% \textbf{(D)}: distance problems; \textbf{(N)}: connectivity problems;
% \textbf{(O)} optimization problems; \textbf{(C)}: centrality problems.} 
%
%\vspace{-2em}
\label{tab:problems}
\vspace{-2em}
\end{table*}

\subsection{Graph Problems and Algorithms}
\label{sec:back_problems}

We next present graph algorithms that have been implemented on FPGAs. 
In the survey, we describe the FPGA designs targeting these algorithms
in~\cref{sec:fpga-algs} (implementations of specific algorithms on the FPGA)
and in~\cref{sec:fpga-frameworks} (implementations within generic graph
processing frameworks on the FPGA).
A summary of the fundamental graph problems, algorithms, and properties
considered in FPGA-related works can be found in Table~\ref{tab:problems}.

\subsubsection{Breadth-First Search (BFS)}

The goal of Breadth-First Search (BFS)~\cite{Cormen:2001:IA:580470} is to visit
each vertex in $G$. BFS starts with a specified \emph{root} vertex~$r$ and
visits all its neighbors $N_r$. Then, it visits all the unvisited neighbors of
$r$'s neighbors, and continues to process each level of neighbors in one
iteration.
During the execution of BFS, the \emph{frontier of vertices} is a data
structure with vertices that have been visited in the previous iteration and
might have edges to unvisited vertices. In the very first iteration, the
frontier consists only of the root~$r$. In each following $i$-th iteration, the
frontier contains vertices with distance~$i$ to the root.  The sequential time
complexity of BFS is $O(n+m)$.

\noindent
\macb{Traditional BFS}
In the traditional BFS formulation, a frontier is implemented with a bag.
At every iteration, vertices are removed in parallel from the bag and all their
unvisited neighbors are inserted into the bag; this process is repeated until the bag
is empty.

\noindent
\macb{Algebraic BFS}
BFS can also be implemented with the MV product over a selected semiring.
For example, for the tropical semiring~\cite{besta2017slimsell}, one updates the vector
of distances from the root by multiplying it at every iteration by
the adjacency matrix.

\subsubsection{Connected Components (CC)}

A connected component is a subgraph of $G$ where any two vertices are connected
by some path.  A \emph{connected} graph consists of only one connected
component. A \emph{disconnected} graph can have several connected components.
In the context of directed graphs, a \emph{strongly connected component} (SCC)
must contain paths from any vertex to any other vertex. In a \emph{weakly
connected component} (WCC), the direction of the edges in a path is not
relevant (i.e., computing WCCs is equivalent to finding CCs when treating the
input directed graph as undirected).
Now, the goal of a CC algorithm is to find all connected components in a
given graph~$G$. 
A simple way to compute CC in linear time is to use BFS and straightforwardly
traverse connected components one by one.
Another established algorithm for CC has been proposed by Shiloach and
Vishkin~\cite{shiloach1980log}. It is based on forming trees among the
connected vertices and then dynamically shortening them using pointer jumping.
In the end, each connected component is represented by one tree consisting of
two levels of vertices: a tree root and its children.
The parallel (under the CRCW PRAM model~\cite{gibbons1989more}) time complexity of
this algorithm is $O(\log n)$.

% A simple way to compute CC in linear time is to use BFS to find one component
% after another: First a random vertex is selected as the starting point for
% the BFS to find all vertices that are connected to it. These vertices make up
% a first connected component $C_1$. If $|C_1| = n$, the graph is shown to
% contain only one component. Otherwise BFS is run again on another previously
% unvisited vertex to find a second connected component.  This is repeated
% until $\sum_i C_i = n$.  Since every vertex is visited only once, but also
% every edge has to be checked, the runtime is $O(n + m)$. 
%
% A very important parallel algorithm for CC has been proposed by Shiloach and
% Vishkin in 1980, also known as the Shiloach-Vishkin algorithm. It is based on
% forming trees among the connected vertices and then dynamically shortening
% them using pointer jumping. 
%
% In essence this problem of CC can be reduced to the problem of merging of
% sets. In the beginning every vertex $i$ belongs to the set $i$. Whenever we
% find an edge that connects two vertices belonging to different sets, the two
% sets are merged. In the end the number of remaining sets equals the number of
% connected components. Such an algorithm has the advantages of being
% parallelizable and having a sequential memory access pattern when accessing
% the graph data which is ideal for achieving high memory throughput. 

\subsubsection{Reachability}

Reachability is a problem strongly related to CC. Namely, it answers the 
question of whether there exists a path between any two vertices. Algorithms
for finding connected components can be used to solve this problem.

\subsubsection{Single-Source Shortest-Paths (SSSP)}

A shortest path between two vertices $v,w$ is a path where either the number of
edges or, in the case of a weighted graph, the sum of all weights in the path,
is the smallest out of all paths between $v$ and $w$. In the Single-Source
Shortest-Paths (SSSP) problem, one finds the shortest path between a given
source vertex and all other vertices in the graph. Two well-known solutions are
Dijkstra's algorithm~\cite{dijkstra1959note} and the Bellman-Ford 
algorithm~\cite{bellman1958routing, ford1956network}. The Bellman-Ford algorithm 
has a sequential time complexity of $O(nm)$ and can be used for graphs with 
negative edge weights, while 
Dijkstra's algorithm, if implemented with a Fibonacci heap, has a better 
sequential time complexity of $O(n \log n + m)$ but cannot handle negative edges. 

\subsubsection{All-Pairs Shortest-Paths (APSP)}

The All-Pairs Shortest-Paths (APSP) problem is to find the shortest paths
between all pairs of vertices in the graph.
% 
% The output of the algorithm should be an n-by-n table, denoting the length of
% the shortest path between any two vertices.  There are two main types of
% algorithms for solving the problem.
%
One solution, called Johnson's algorithm, is to use the SSSP algorithms such as
Dijkstra and Bellman-Ford~\cite{johnson1977efficient}. In case of unweighted
graphs, the algorithm can be further reduced to BFS. The worst-case sequential
runtime is $O(n^2 \log n + nm)$. Johnson's algorithm for weighted graphs
requires a Fibonacci heap, which may be difficult to implement. 
Another solution is the Floyd-Warshall algorithm~\cite{floyd1962algorithm},
which has the $O(n^3)$ sequential time complexity and is based on dynamic
programming.

% There are two papers that solve the APSP problem on the FPGA, one of them 
% uses the Floyd-Warshall algorithm \cite{bondhugula:APSP_FPGA} and 
% the other one is based on BFS \cite{betkaoui:APSP_FPGA}. 

\subsubsection{Minimum Spanning Tree (MST)}
A spanning tree of a graph is defined as a tree subgraph that includes all the vertices of the graph and a subset of the edges with minimal size. An MST is thus a spanning tree where the edge weight sum is minimal. There exist several algorithms to find the MST of a graph, notably Boruvka's algorithm~\cite{boruuvka1926jistem}, which runs in parallel at a time complexity of $O\left(\log n\right)$; the sequential Prim's Algorithm~\cite{prim1957shortest} with $O\left(m+n\log n\right)$ complexity (using a Fibonacci heap); and Kruskal's Algorithm~\cite{kruskal1956shortest} with $O\left(m \log n \right)$ time complexity.

\subsubsection{PageRank (PR)}

PageRank (PR)~\cite{page1999pagerank} is an iterative centrality algorithm that
obtains the rank $r(v)$ of each vertex $v$: 

$$
r(v) = \frac{1 - f}{n} + \sum_{w \in N_v} \frac{f \cdot r(w)}{d_v}
$$

\noindent
where $f$ is a parameter called the damp factor~\cite{page1999pagerank}. PR is
used to rank websites.  Intuitively, a vertex is deemed ``important'' (has a
high rank) if it is being referenced by other high rank vertices.

The papers covered in this survey implement the traditional iterative PR
algorithm where, in each iteration, all edges are processed and the PR of every
vertex is recomputed according to the above equation.  Usually the maximum
number of iterations is set, the algorithm halts if the maximum difference
between the ranks of a vertex in two iterations converges below a given
threshold. 

% The algorithm performs several iterations, where in each one the page rank
% value of every vertex is recomputed as the sum of its neighbors' page rank
% values, multiplied by some dampening factor. 

% \subsubsection{Minimum Spanning Tree (MST)}
% NOTE: not really relevant
% The goal of the Minimum Spanning Tree (MST) problem is to derive a spanning
% tree of $G$ with the lowest sum of the included edge weights. The classical
% algorithms are Prim \cite{prim1957shortest}, 
% kruskal \cite{kruskal1956shortest}, and boruvka \cite{nevsetvril2001otakar}.

\subsubsection{Graphlet Counting (GC), Triangle Counting (TC)}

Graphlets are small connected induced subgraphs. An example graphlet is a
triangle: a cycle with three vertices. The problem of counting
graphlets (GC) is to count the number of different graphlets in a graph.
There exist many algorithms for counting graphlets of different sizes.  For
example, TC can be solved in $O(m^{3/2})$
time~\cite{schank2007algorithmic}.

% The naive graphlet counting algorithm goes as follows: 
% 
% \begin{itemize}
%   \item For each vertex $a$ of the graph, iterate through all of its neighbors $b$ and then 
%   through all of $b$'s neighbors $c$ (where $a \neq c$) and check whether $a, b$ and $c$ form a graphlet of size 3. 
%
%   \item Iterate through all of $c$'s neighbors $d$ and check whether 
%   $a, b, c$ and $d$ are a graphlet of size 4. 
%
%   \item Scan all of $d$'s neighbors $e$ and check 
%   whether $a, b, c, d$ and $e$ are a graphlet of size 5. 
% \end{itemize}
% 
% The algorithm consists of five nested loops and thus has a complexity of $O(n^5)$. The major
% problem for large sparse graphs lies in the irregular and unpredictable memory 
% access pattern and the performance is thus dominated by memory latencies. 

\subsubsection{Betweenness Centrality (BC)}

Centrality measures of vertices in a graph determine the ``importance'' of each vertex . One such measure is Betweenness Centrality~\cite{newman2005measure}, which is defined by the ratio of shortest paths in the graph that pass through a given vertex . More formally:
$$
BC(v) = \sum_{\substack{u,w\in V\\ u,w \neq v}}{\frac{P_v\left(u,w\right)}{P\left(u,w\right)}},
$$
where $P(u,w)$ indicates the number of shortest paths between $u$ and $w$ and $P_v$ is the number of shortest paths that pass through $v$. To compute BC for every vertex, one can use the Brandes algorithm~\cite{brandes2001faster} in parallel~\cite{solomonik2017scaling}, which exhibits a total work of $O\left(nm\right)$.

\subsubsection{Maximum Matching (MM)}
A matching is defined to be a set of edges $E'\subseteq E$, where every vertex in the pairs of $E'$ is unique, i.e., edges do not share vertices. Maximum Cardinality Matching and Maximum Weighted Matching (MWM) are commonly computed types of matchings, where the former is a matching that maximizes $|E'|$ and the latter (only applicable to weighted graphs) maximizes the sum of the edge weights in $E'$. MWM can be computed exactly using the Blossom Algorithm~\cite{edmonds1965paths}, which is inherently sequential.

% \subsubsection{Minimum Vertex Cover (VCover)}
% \talbn{not sure about this}

\subsubsection{Graph-Related Applications}

In the same way that some graph algorithms (e.g., BFS) can be implemented with
linear algebra operators, many applications outside of graph theory formulate
problems as graphs in order to benefit from increased performance. One such
example is Deep Neural Networks and the Stochastic Gradient Descent algorithm:
in the former, sparsely-connected layers of neurons can be represented as a
graph that should be traversed~\cite{cambriconx}; whereas in the latter, the
algorithm itself creates dependencies that can be modeled as a fine-grained
graph and scheduled dynamically~\cite{kaleem15sgd}.

Another graph-related application considered in the surveyed works is stereo
matching~\cite{szeliski2008comparative}, where one accepts a pair of stereo
images and outputs the disparity map with depth information of all pixels.
This problem is solved with the Tree-Reweighted Message Passing (TRW-S)
algorithm~\cite{szeliski2008comparative}.

Finally, one work considers spreading activation~\cite{liu2004conceptnet}.
Here, a given graph is traversed (starting from a selected subset of vertices)
in a manner similar to that of a neural network: certain values called
activities are propagated through the graph, updating properties of vertices.

\subsubsection{Challenges in Graph Processing for FPGAs}

Most challenges for efficient graph processing on FPGAs are similar across
different algorithms. BFS, the algorithm considered most often,
exhibits irregular memory access patterns because it is hard to predict
which vertices will be accessed in the next iteration, before
iterating through the neighborhood of vertices currently in the frontier.
Vertices with small distances between them in $G$ are not necessarily close to
one another in memory. Furthermore, most vertices in graphs used in today's
computations have small neighborhoods and thus prefetching subsequent memory
blocks does not always improve performance.

\subsection{Graph Programming Paradigms, Models, and Techniques}
\label{sec:back_models}

We also present graph programming models used in the surveyed works. 
A detailed description can be found in work by Kalavri et al.~\cite{kalavri2018high}.
% 
% Many of the
% papers we discuss in this survey argue that one or the other programming model
% is more suitable for FPGA implementations. Choosing the right programming model
% can be key to optimizing memory access patterns and thus improving performance. 

\subsubsection{Vertex-Centric Model}

In the vertex-centric model~\cite{khan2016query, malewicz2010pregel},
the programmer expresses a graph algorithm from the perspective of
vertices.
%
% The vertex-centric model~\cite{khan2016query, malewicz2010pregel} is often 
% used to develop graph algorithms.
%
One programs an algorithm by developing a (usually small)
routine that is executed \emph{for each vertex in the graph concurrently}. In
this routine, one usually has access to the neighbors of a given vertex.
Such as approach can lead to many random memory accesses as neighboring
vertices may be stored in different regions of the memory. Still, it is often
used because many important algorithms such as BFS or PageRank can easily be
implemented in this model.

% Note  mentioning Pregel \cite{malewicz2010pregel} may not be necessary

\subsubsection{Edge-Centric Streaming Model}

In the edge-centric model~\cite{roy2013x}, edges are streamed to 
the processor in the
order in which they are stored in the graph data structure. An edge consists of
the labels of the two connected vertices and optionally the edge weight.  The
processor processes edges one by one and, if necessary, it updates some
associated data structures.
This way of accessing the graph has major advantages because of its sequential
memory access pattern, which improves spatial locality.
A disadvantage of this approach is the restriction on the order of loading and thus processing 
edges.  This makes the model less suitable for certain algorithms, for example
BFS or SSSP. For example, the edge-centric BFS requires several passes over the
edges, with each pass only processing those edges that are currently in the
frontier. This takes $O(Dm)$ time, a factor of $D$ more than when using the
traditional BFS variant with $O(m)$ time~\cite{besta2017push}.

% There is a class of algorithms called graph-streaming algorithms, which is
% based on processing a graph as a stream of edges. FPGAs are suitable for
% dataflow streaming applications, making use of the very high internal data
% bandwidth between different components on the FPGA
% \cite{neuendorffer2008streaming}.  However many graph algorithms by nature
% would not require very large pipelines since they are mostly data-intense and
% not very compute-heavy. 

\subsubsection{Gather-Apply-Scatter Model (GAS)}

Gather-Apply-Scatter (GAS)~\cite{low2010graphlab, kyrola2012graphchi} 
is similar to the vertex-centric
approach. It also offers the vertex-centric view on graph algorithms. However,
it additionally splits each iteration into three parts: gather, apply, and
scatter. In the \textbf{gather} phase, a vertex collects information about its
neighboring vertices or edges and optionally reduces them to a single value
$\sigma$. In the \textbf{apply} stage, the state of the vertex is updated
based on the previously computed $\sigma$, and possibly the properties of the vertex
neighbors. Finally, in the \textbf{scatter} phase, each vertex propagates its
new state to its neighbors. This value will then again be collected in the
gather phase of the next iteration. 
These three phases can be implemented as individual components and for example
connected as a pipeline system or in a network of distributed components.

% There are also some variations, such as \textit{scatter-gather}, which is
% basically the same model but with a different order.  The GAS model can be
% implemented in both the vertex-centric or the edge-centric models. 
% 
% In an edge-centric model, in the scatter phase the algorithm would create
% messages to the destination vertices of each edge, whereas in the gather
% phase these messages would be collected and reduced to update the state of
% those vertices.
% 
% The model is often used in distributed graph algorithms and in GPU
% implementations because its perfect for distributing work among different
% compute nodes.  Some examples are VertexAPI2 (Elsen and Vaidyanathan 2013),
% MapGraph (Fu et al.  2014), and CuSha (Khorasani et al. 2014). 
% 
% This model should not be confused with a type of memory addressing that is
% also called 'gather-scatter'. 

\subsubsection{Bulk-Synchronous Parallel (BSP) Model}

Bulk-Synchronous Parallel (BSP)~\cite{valiant1990bridging}
is a model for designing and analyzing general (i.e., \emph{not} specifically
graph-related) algorithms.  Each iteration in the model is called a
\textbf{superstep}. After each superstep, parallel processes are synchronized
using a barrier. Similarly to the GAS model, each iteration is divided into
three phases. In first phase, each process conducts any required local
computation. In the next phase, the processes send and receive messages.
Finally, a barrier synchronization guarantees that the next super-step only
begins after all local computations in the current super-step are finished and
all messages in this super-step are exchanged.
BSP is frequently used to model and analyze distributed graph
algorithms~\cite{gianinazzi2018communication, ediger2013investigating,
caceres1997efficient}.

% This model is slightly less strict than GAP since the tasks that are computed
% locally can be arbitrarily as long as they are limited on the vertices
% belonging to that process. It can thus be used to implement distributed
% algorithms that partition the graph. 
% 
% There is no strict synchronization between the first and the second phase
% which means that one process can start with the communcation phase while
% other processes are still in the computing phase.  However the barrier
% between the super-steps is enforces a synchronization betweeen all the nodes
% and thus the system progresses only when the slowest node has finished. 

% This model is slightly less strict than GAP since the tasks that are computed
% locally can be arbitrarily as long as they are limited on the vertices
% belonging to that process. It can thus be used to implement distributed
% algorithms that partition the graph. 
% 
% There is no strict synchronization between the first and the second phase
% which means that one process can start with the communcation phase while
% other processes are still in the computing phase.  However the barrier
% between the super-steps is enforces a synchronization betweeen all the nodes
% and thus the system progresses only when the slowest node has finished. 

\subsubsection{Asynchronous Execution}

While BSP imposes a strict global barrier after each iteration, in an
asynchronous execution model~\cite{low2010graphlab}, some processes can advance
to the next iteration even before others have finished the previous iteration.
In the context of iterative graph algorithms such as PageRank or
Shiloach-Vishkin, this enables some vertices to propagate their associated
values (e.g., their ranks) to their neighbors more often than others, i.e, not
just once per iteration. This can accelerate convergence, but it also requires
more complex synchronization mechanisms than in algorithms that use
synchronous models.

\subsubsection{MapReduce (MR)}

MapReduce (MR)~\cite{dean2008mapreduce} is a well-known programming model 
for processing data
sets in a parallel, distributed setting. An algorithm based on the MapReduce
model is usually composed of several iterations, and each iteration consists of
three phases: map, shuffle, reduce.  In the \textbf{map} phase, a certain map
function is applied to every input value and an intermediary result is
generated.  Next, in the \textbf{shuffle} phase, compute nodes can redistribute
the outcomes of the map phase.  In the final \textbf{reduce} phase, a certain
reduction function is performed by each node on the data received in the
shuffle phase.
Such MapReduce iterations can be used to expressed some graph
algorithms~\cite{cohen2009graph}.

% A MapReduce program is defined as a single iteration of the above steps. Many
% graphs however consist of more than one iteration. Thus the MapReduce model
% is not necessarily a good fit for graph computing. Another requirement for
% distributed graph computing is message passing.  Worker nodes that compute
% different partitions of a graph need to pass information to one another
% between iterations. 

\subsubsection{Substream-Centric}

The substream-centric approach~\cite{besta2019substream} is a paradigm designed
to compute semi-streaming graph algorithms efficiently. In substream-centric
algorithms, an input stream of data is divided into substreams, processed
independently to increase parallelism while lowering communication costs.
Specifically, the semi-streaming model assumes that the input is a sequence of
edges, which can be accessed only sequentially, as a stream. The main memory
(can be randomly accessed) is assumed to be of size
$O\left(n~~\text{polylog}~n\right)$.  Usually, only one pass over the input
stream is allowed, but some algorithms assume a small (usually constant or
logarithmic) number of passes. The approach then divides the incoming stream of
edges into substreams, processes each substream independently, and merges these
results to form the final algorithm outcome.

%\subsubsection{General Techniques}
%\maciej{Describe Dynamic Programming, Transactional Memory}

\subsection{FPGA Architecture and Terminology}

\begin{figure*}[t]
\centering
\vspace{-1em}
\includegraphics[width=1.0\textwidth]{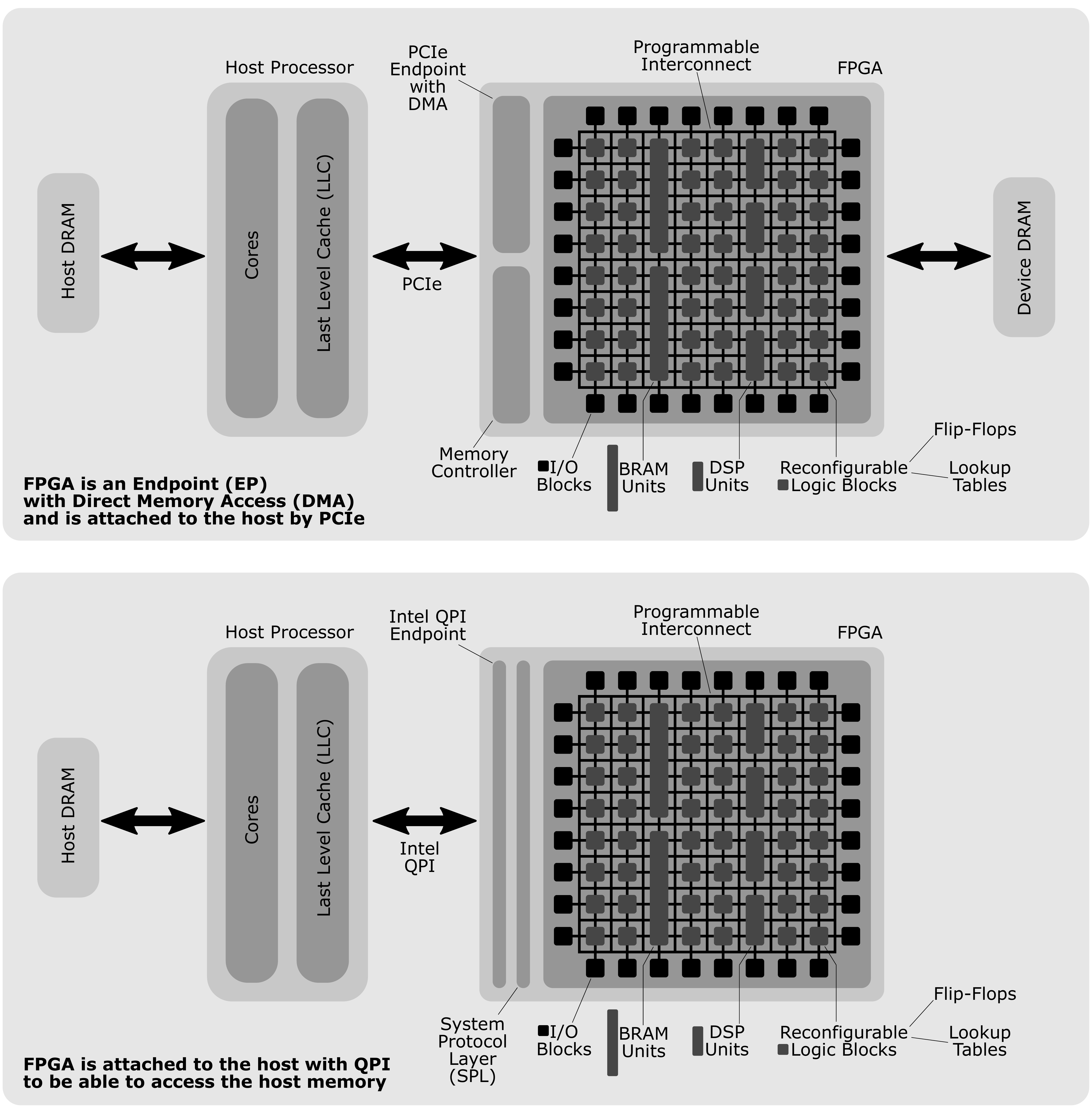}
\vspace{-2em}
\caption{Illustration of an FPGA and of two possible hybrid FPGA--CPU computation systems.}
\label{fig:fpga}
\end{figure*}

Field-programmable gate arrays (FPGAs) are reconfigurable computing devices that
contain a large number of programmable units that can be used to solve specific
computational problems (see Figure~\ref{fig:fpga}). These logic units include
lookup tables (LUTs) to implement combinatorial logic, flip-flops to implement
registers, and a programmable interconnect. FPGAs often also provide more
specialized units, such as block RAM (BRAM) for bulk storage, and DSP units for
accelerating common arithmetic operations. 

In contrast to application-specific integrated circuits (ASICs), which are only
``configured'' once during the manufacturing process, FPGAs can be reconfigured
as often as needed. This allows improving and changing the architecture,
applying bug-fixes, or using FPGAs to rapidly prototype hardware designs, which
can later be manufactured as ASICs. FPGAs additionally allow reconfiguration on
the fly to solve different tasks~\cite{hauck2010reconfigurable}.

While CPUs and GPUs are instruction-driven, FPGA designs are usually
data-driven, which means that an FPGA can process data directly without having
to first decode instructions, and does not access a centralized register file
or any cache hierarchy. This is usually more power efficient, as instruction
decoding, register file lookup, and cache lookup account for the majority of
power consumed on instruction-based architectures~\cite{horowitz_energy}.

The reconfigurability comes with the cost of a lowered frequency, usually about
3-10 times lower than that of CPUs, and with less specialized
components, e.g., floating point operations are often not native operations, and
must be implemented with general purpose logic. Still, carefully engineered
FPGA designs can outperform CPU implementations, by exploiting massive
parallelism, typically in the form of deep pipelines. As long as there are no
feedback data dependencies between iterations of an iterative algorithm,
arbitrarily complex computations can be implemented as a pipeline that can
produce one result per cycle.
Application-specific instructions that are not a part of a CPU instruction set,
e.g., a novel hash function, can be implemented on an FPGA to deliver a
result every cycle, whereas a CPU implementation would potentially require many
CPU instructions.

When FPGA performance models are discussed, we denote the bandwidth between the FPGA
and DRAM as $B_{DRAM}$. The bandwidth of a single BRAM module is denoted as
$B_{BRAM}$.

\subsubsection{FPGA Programming Languages}

The traditional way to program an FPGA is to use a hardware description language
(HDL) such as Verilog, VHDL, or SystemC. These languages describe the
cycle-by-cycle behavior of hardware at the register transfer level (RTL), which
can then be synthesized to the underlying hardware resources and used to
configure an FPGA.  The low-level nature of these languages means that they lack
many of the high-level concepts that are found in software programming
languages.  An alternative is to generate HDL using a high-level synthesis (HLS)
tool, where the hardware description is derived from a more
high-level imperative description. Many HLS tools exist~\cite{hls_survey}, and
most are based on C, C++ or OpenCL, using directives to aid the developer in
expressing architectural features, such as pipelines, hardware replication, and
fast memory allocation, in the generated RTL code.
Other approaches include the Scala-based Chisel~\cite{chisel} that offers a
more productive environment for RTL development, and
the commercial MaxCompiler~\cite{maxeler}, which compiles to hardware from a
dataflow-oriented Java-based language. 

% PAM-Bloks II
% SystemVerilog
% Migen

\subsubsection{Coarsening of FPGA Features}

Recent development has seen increasing specialization and diversification in
FPGA architectures. Intel's Arria~10 and Stratix~10 families of FPGAs offer
native 32-bit floating point (FP32) units, which greatly reduces the area usage
of these operations (although 64-bit floating point is still costly), and
simplifies certain patterns by supporting native accumulation of FP32 data.
Stratix~10 FPGAs also expose ``HyperFlex''~\cite{hyperflex} registers, a new
family of dedicated routing registers aimed at improving frequency results, in
order to narrow the gap to CPU and GPU clock rates. Xilinx UltraScale+ devices
add a new class of on-chip RAM called UltraRAM~\cite{ultraram}, that expose
access ports of similar width to traditional block RAM, but have larger
capacity, allowing large amounts of memory to be stored on the chip without
requiring as many individual RAM blocks to be combined.  Finally, the
Versal~\cite{versal} family of Xilinx devices puts the FPGA on the same chip as
an array of ``AI engines'', capable of performing more traditional SIMD-style
arithmetic operations, adding to the compute potential in a hybrid ASIC/FPGA
fashion. Common for these trends is a \emph{coarsening} of components,
sacrificing some flexibility for more raw performance and on-chip memory
bandwidth for suitable workloads.

\subsubsection{Integration with Hybrid Memory Cubes}
\label{sec:back-HMC}

A number of surveyed works relies on using the combination of FPGAs and the
Hybrid Memory Cube (HMC) technology~\cite{pawlowski2011hybrid}. HMC dramatically improves the
bandwidth of DRAM. An HMC unit consists of multiple DRAM dies that are connected
using the \emph{through-silicon-via} (TSV) technology. For example, it offers a
8--10$\times$ bandwidth improvement over DDR4 and
is optimized for parallel random memory access~\cite{pawlowski2011hybrid}. Compared to
traditional DRAM, HMC has much smaller page sizes (16B), which offers more
performance for random memory accesses.
% 
% In contrast, large page sizes are more efficient with high data locality but
% consume energy unnecessarily when data locality is poor. This is known as the
% over-fetching problem.
%
Furthermore, HMC implements near memory computing in the form of locking and
read-modify-write operations that are computed directly by the HMC unit instead
of the CPU. It is even possible to atomically modify individual bits without
having to first read the corresponding bytes.
%
%This is beneficial for manipulating bitmaps and is used by Zhang et al. to
%efficiently update the frontier in their BFS implementation.

\section{Taxonomy}

% and the space of the considered schemes.

%% make a table summarizeing the papers

% The use of FPGAs for graph processing is a rather recent development and there 
% are only a few dozen papers about the topic. 

\begin{figure*}%[!h]
\centering
\includegraphics[width=0.75\textwidth]{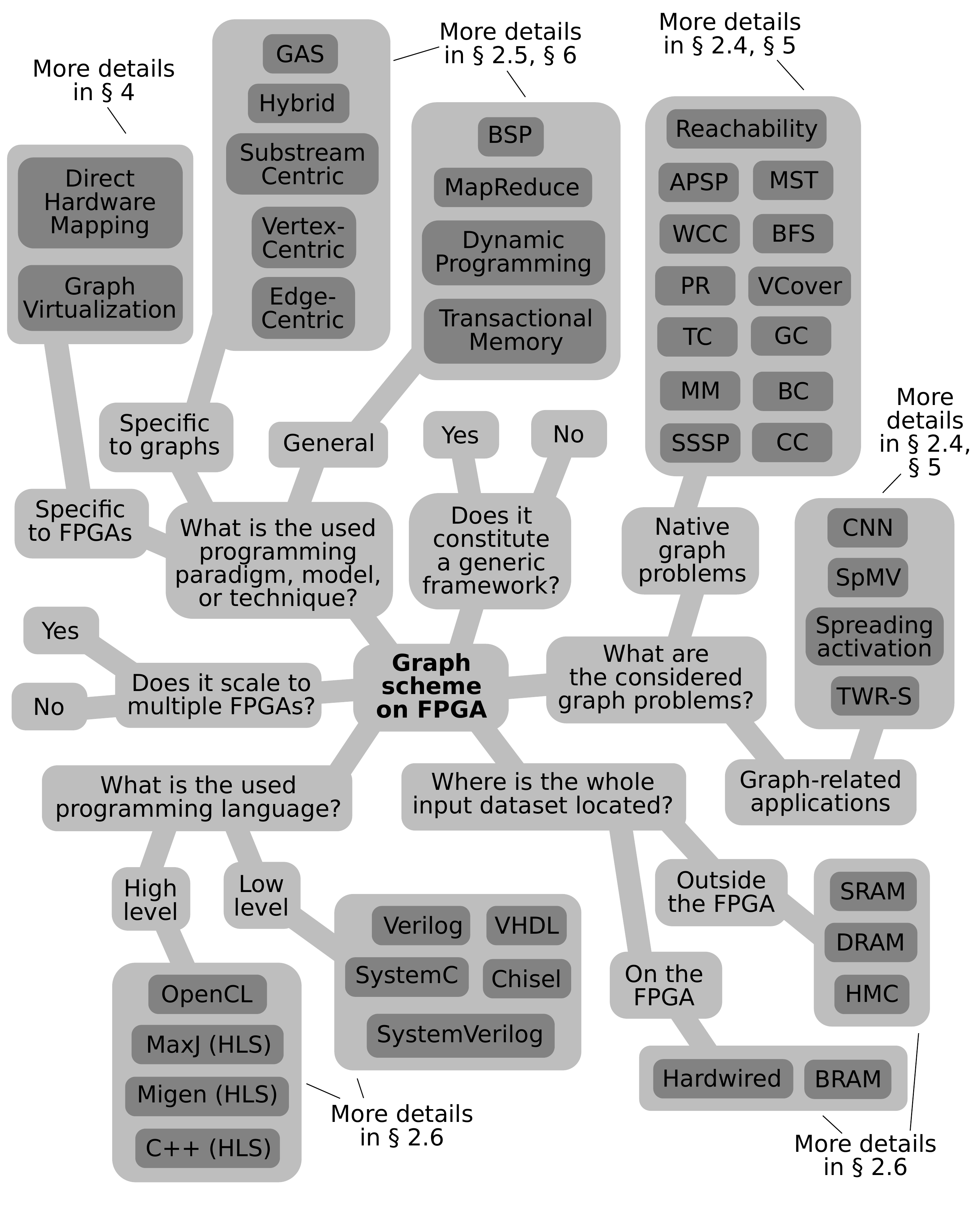}
\vspace{-1em}
\caption{The categorization of the considered domains of graph processing on FPGAs.
All the categories are gathered in a form of a tree. The tree root in the centre
represents all graph processing schemes implemented on FPGAs.
The children of the tree root correspond to various general criteria that
can be used to categorize FPGA graph processing schemes, e.g., whether or not
a given FPGA scheme constitutes a generic graph processing framework
or whether it is an implementation of a particular algorithm.}
\vspace{-1em}
\label{fig:categorization}
\end{figure*}

In this section, we describe how we categorize the surveyed work.
We summarize the most relevant papers in Table~\ref{tab:papers}. We group
separately generic graph processing frameworks and specific algorithm
implementations.  Each group is sorted chronologically. Selected columns in
this table constitute criteria used to categorize the surveyed FPGA works, see
also Figure~\ref{fig:categorization}.

% In this survey, we consider any work that deals with processing graph
% algorithms on FPGA devices. However, we also mention works that target other
% accelerator hardware but present ideas relevant for FPGAs. 
%
The first such criterion is \textbf{generality}, i.e., whether a given FPGA
scheme is focused on a particular graph problem or whether it constitutes a
generic framework that facilitates implementing different graph algorithms. 
% 
% Some works describe complete frameworks that enable users to express their
% own graph processing application by implementing some key functionality and
% then compile the solution for any given FPGA device. 
%
Another criterion is a used \textbf{graph programming paradigm, model, or
technique}. We describe the general paradigms, models, and techniques in detail
in~\cref{sec:back_models}. However, certain techniques for graph processing are
specific to FPGAs; we cover such techniques separately
in~\cref{sec:fpga-techniques}. Note that many implementations are not based on
any particular paradigm or model and they do not use any particular general
technique; we denote such works with ``None''.
% 
% Some works show that a vertex-centric model is suitable for FPGAs, whereas
% others adopt the edge-centric model with the goal to improve spatial
% locality. Certain designs are hybrid, such as the one proposed by Zhou and
% Prasanna~\cite{zhou2017accelerating}; they can switch between the two models.
% Other utilized models are GAS, BSP, and Asynchronous Execution.

We also distinguish between works that target a single FPGA and ones that scale
to \textbf{multiple FPGAs}. Finally, we consider the used \textbf{programming
language} and the \textbf{storage location} of the \emph{whole} processed graph
datasets.  In the latter, ``DRAM'', ``SRAM'', or ``HMC'' indicates that the
input dataset is located in DRAM, SRAM, or HMC, and it is streamed in and out
of the FPGA during processing (i.e., only a part of the input dataset is stored
in BRAM at a time). Contrarily, ``BRAM'' indicates that the whole dataset is
assumed to be located in BRAM.  ``Hardwired'' indicates that the input dataset
is encoded in the FPGA reconfigurable logic.

\begin{table*}
\vspace{-1em}
\centering
 \setlength{\tabcolsep}{1pt}
\scriptsize
%\footnotesize
%\ssmall
\sf
\begin{tabular}{@{}llllllllrr@{}}
\toprule
%
% \textbf{Name} & \textbf{Reference} & \makecell[c]{\textbf{Scalable to}\\\textbf{Multiple FPGAs}} & \textbf{Supported Algorithms} & \textbf{Programming Model} & \textbf{Generic Framework}& \textbf{Max Graph Size} & \textbf{Max Graph Size} \\
	\makecell[c]{\textbf{Reference}\\\textbf{(scheme name)}} & \makecell[c]{\textbf{Venue}} & \makecell[c]{\textbf{Generic}\\\textbf{Design}$^1$} & \makecell[c]{\textbf{Considered}\\\textbf{Problems$^2$~(\cref{sec:back_problems})}} & \makecell[c]{\textbf{Programming Model}\\\textbf{or Technique$^4$~(\cref{sec:back_models})}}
  & \makecell[c]{\textbf{Used}\\\textbf{Language}}
	& \makecell[c]{\textbf{Multi}\\\textbf{FPGAs}$^4$}
	& \makecell[c]{\textbf{Input} \\\textbf{Location}$^5$} & $n^{\text{\textdagger}}$ & $m^{\text{\textdagger}}$ \\
% & \makecell[c]{\textbf{Achieved}\\\textbf{MTEPS}$^6$}\\
%
\midrule
\makecell[l]{Kapre \cite{kapre2006graphstep}\\\textbf{(GraphStep)}}             & FCCM'06       & \faThumbsOUp & \makecell[l]{spreading\\activation$^*$~\cite{liu2004conceptnet}}   & BSP     &unsp. & \faThumbsOUp & BRAM       & 220k  	&  550k    \\ % & 0.7    \\
\makecell[l]{Weisz \cite{weisz:GraphGen}\\\textbf{(GraphGen)}}                     & FCCM'14       & \faThumbsOUp & \makecell[l]{TRW-S$^*$,\\CNN$^*$~\cite{sun2003stereo}}  & Vertex-Centric   &unsp. &\faThumbsDown & DRAM  & 110k    & 221k  \\ %  & 9.9   \\
\makecell[l]{Kapre \cite{kapre:custom_graph_FPGA}\\\textbf{(GraphSoC)}}            & ASAP'15       & \faThumbsOUp & SpMV                        & Vertex-Centric, BSP  & C++ (HLS)  & \faThumbsOUp & BRAM  & 17k     & 126k  \\ %  &unsp.   \\
\makecell[l]{Dai \cite{dai:fpgp}\\\textbf{(FPGP)}}                             & FPGA'16       & \faThumbsOUp  & BFS                         & None                      &unsp. & \faThumbsOUp & DRAM  & 41.6M   & 1.4B \\ %   & 12  \\
\makecell[l]{Oguntebi \cite{oguntebi:GraphOps}\\\textbf{(GraphOps)}}               & FPGA'16       & \faThumbsOUp & \makecell[l]{BFS, SpMV, PR,\\Vertex Cover}      & None                      & MaxJ (HLS)  &\faThumbsDown & BRAM  & 16M     & 128M  \\ %  & 200  \\
Zhou \cite{zhou2016high}         & FCCM'16       & \faThumbsOUp & \makecell[l]{SSSP, WCC, MST}                & Edge-Centric          &unsp. &\faThumbsDown & DRAM  & 4.7M    & 65.8M \\ %  & 730   \\
\makecell[l]{Engelhardt \cite{engelhardt2016gravf}\\\textbf{(GraVF)}}           & FPL'16        & \faThumbsOUp & \makecell[l]{BFS, PR, SSSP, CC}              & Vertex-Centric        & \makecell[l]{Migen\\(HLS)} &\faThumbsDown & BRAM  & 128k    & 512k \\ %   & 3000 \\
\makecell[l]{Dai \cite{dai:foregraph}\\\textbf{(ForeGraph)}}                        & FPGA'17       & \faThumbsOUp & \makecell[l]{PR, BFS, WCC}                &  None                     &unsp. & \faThumbsOUp & DRAM  & 41.6M  & 1.4B  \\ %  & 1000  \\
Zhou \cite{zhou2017accelerating} & SBAC-PAD'17   & \faThumbsOUp &  BFS, SSSP            & \makecell[l]{Hybrid (Vertex-\\and Edge-Centric)}  &unsp. &\faThumbsDown & DRAM  & 10M    & 160M   \\ % & 670    \\
Ma \cite{ma2017fpga}                            & FPGA'17       & \faThumbsOUp & \makecell[l]{BFS, SSSP, CC,\\TC, BC}       & \makecell[l]{Transactional\\Memory~\cite{herlihy1993transactional, besta2015accelerating}}  & \makecell[l]{System-\\Verilog}  & \faThumbsOUp & DRAM  & 24M    & 58M   \\ %  &unsp.    \\
\makecell[l]{Lee \cite{lee2017extrav}\\\textbf{(ExtraV)}}                        & FPGA'17       & \faThumbsOUp & \makecell[l]{BFS, PR, CC,\\AT$^*$~\cite{hong2014simplifying}} & Graph Virtualization  & C++ (HLS)  &\faThumbsDown & DRAM  & 124M   & 1.8B  \\ %  &unsp. \\
Zhou \cite{zhou2018framework}     & CF'18         & \faThumbsOUp & SpMV, PR                    & Edge-Centric, GAS    &unsp.      &\faThumbsDown & DRAM  & 41.6M  & 1.4B  \\ %  & 2250  \\
Yang \cite{yang2018efficient}                   & report (2018)        & \faThumbsOUp & BFS, PR, WCC                & None                      & OpenCL  &\faThumbsDown &   & 4.85M  & 69M   \\ %  & 175  \\
Yao \cite{yao2018efficient}                     & report (2018)        & \faThumbsOUp & BFS, PR, WCC                &      None                 &unsp.      &\faThumbsDown & BRAM  & 4.85M  & 69M  \\ %   & 3200  \\
\midrule
Babb \cite{babb1996solving}                     & report (1996)        & \faThumbsDown & SSSP                                  & None      & Verilog  & \faThumbsOUp & Hardwired  & 512   	&  2051   \\ %  &unsp.     \\
Dandalis \cite{dandalis1999domain}              & report (1999)        & \faThumbsDown & SSSP                                  & None      &unsp. & \faThumbsOUp & Hardwired  & 2048  	&  32k   \\ %  &unsp.     \\
Tommiska \cite{tommiska2001dijkstra}            & report (2001)        & \faThumbsDown & SSSP                                  & None                      & VHDL  & \faThumbsDown  & BRAM        & 64    	&  4096 \\ %   &unsp.     \\
Mencer \cite{mencer2002hagar}                   & FPL'02        & \faThumbsDown & \makecell[l]{Reachability,\\SSSP}                    & None      & \makecell[l]{PAM-\\-Blox II}  & \faThumbsDown  & \makecell[l]{Hardwired\\(3-state\\buffers)}  & 88    	&  7744 \\% &unsp.    \\
Bondhugula \cite{bondhugula:APSP_FPGA}          & IPDPS'06      & \faThumbsDown & APSP                                  & \makecell[l]{Dynamic Program.}   &unsp. &\faThumbsDown & DRAM  &unsp. 	& \\%	&unsp.     \\
Sridharan\cite{sridharan2009hardware}           & TENCON'09     & \faThumbsDown & SSSP                                  &  None                     & VHDL  &\faThumbsDown & BRAM  & 64     	& 88    \\%  &unsp.     \\
Wang \cite{wang2010message}                     & ICFTP'10      & \faThumbsDown & BFS                                   &  None                     & SystemC  &     & DRAM  & 65.5k  	& 1M      \\% & 795   \\
Betkaoui \cite{betkaoui2011framework}           & FTP'11        & \faThumbsDown & GC                                    & Vertex-Centric        & Verilog  & \faThumbsOUp & DRAM  & 300k 	& 3M     \\% unsp.     \\
Jagadeesh \cite{jagadeesh2011field}             & report (2011)        & \faThumbsDown & SSSP                                  & None      & VHDL  &\faThumbsDown & Hardwired  & 128    & 466    \\ % &unsp.\\
Betkaoui \cite{betkaoui:APSP_FPGA}              & FPL'12        & \faThumbsDown & APSP                                  & Vertex-Centric        & Verilog  & \faThumbsOUp & $\approx $ DRAM  & 38k     & 72M    \\ % &       \\
Betkaoui\cite{betkaoui2012reconfigurable}       & ASAP'12       & \faThumbsDown & BFS                                   & Vertex-Centric        & Verilog  & \faThumbsOUp & DRAM  & 16.8M  & 1.1B    \\ % & 720  \\
\makecell[l]{Attia~\cite{attia2014cygraph}\\\textbf{(CyGraph)}}                   & IPDPS'14      & \faThumbsDown & BFS                                   & Vertex-Centric        & VHDL  & \faThumbsOUp & DRAM  & 8.4M    & 536M  \\ %  & 550   \\
Ni \cite{ni2014parallel}                        & report (2014)        & \faThumbsDown & BFS                                   &  None                     & Verilog  &\faThumbsDown & \makecell[l]{DRAM,\\SRAM}  & 16M     & 512M  \\ %  & 950   \\
Zhou \cite{zhou2015sssp}          & IPDPS'15      & \faThumbsDown & SSSP                                  &     None                  &unsp. &\faThumbsDown & DRAM  & 1M      &unsp.    \\ %  & 1600  \\
Zhou \cite{zhou2015pagerank}      & ReConFig'15   & \faThumbsDown & PR                                    & Edge-Centric          &unsp. &\faThumbsDown & DRAM  & 2.4M    & 5M    \\ %  &       \\
Umuroglu \cite{umuroglu:hybrid_bfs_FPGA}        & FPL'15        & \faThumbsDown & BFS                                   &     None                  & Chisel  & & $\approx$ DRAM     & 2.1M    & 65M  \\ %   & 172   \\
%Kapre \cite{kapre:SoCs_FPGA}                    & report (2015)        &unsp.                                    & Vertex-Centric / BSP  &   &\faThumbsDown &   & 32M     & 32M     & 91-95 \\
% Ham \cite{ham2016graphicionado}                 & MICRO'16      & generic (PR, BFS, SSSP, CF$^{\text{\textdagger}}$)   & Vertex-Centric   & Chisel  &\faThumbsDown & eDRAM & 61.5M  & 1.4b    & 2500      \\
%Ozdal \cite{ozdal2016energy}                    & ISCA'16       & \faThumbsOUp & \makecell[l]{PR, LBP, SGD, SSSP}          & Async. Execution & SystemC & \faThumbsOUp & DRAM & 67M    & 1B      &        \\
Lei \cite{lei2016fpga}                          & report (2016)      & \faThumbsDown  & SSSP                                  & None        &unsp. &\faThumbsDown & DRAM  & 23.9M  & 58.2M \\ %  & \\
Zhang \cite{zhang:graph_FPGA}                   & FPGA'17       & \faThumbsDown & BFS                                   & MapReduce             &unsp. &\faThumbsDown & HMC & 33.6M  & 536.9M \\ %  & 166.2 \\
Zhang \cite{zhang2018degree}                    & FPGA'18       & \faThumbsDown & BFS                                   & None                      &unsp.      &     & HMC  &                      \\ %   & 45800      \\
Kohram \cite{khoram2018accelerating}  & FPGA'18       & \faThumbsDown & BFS                                   &        None               &unsp.      &\faThumbsDown  & HMC &                      \\ %   &    \\
Besta~\cite{besta2019substream}  & FPGA'19 & \faThumbsDown & MM & Substream-Centric & Verilog & \faThumbsDown & DRAM & 4.8M & 117M \\ % &unsp.\\
% TuNao is not FPGA related
% J.Zhou \cite{j_zhou:TuNao}                      & CCGrid'17     & generic (PR, BFS, SSSP, CC, ...)      & GAS Model             &\faThumbsDown & |V|=4.2m,   |E|=101m    &       \\
% Weisz \cite{weisz:GraphGen_for_CoRAM}           & (2013)        & Stereo-Matching   & Vertex-Centric           &\faThumbsDown & |V|=100k,   |E|=200k &        \\
% Horawalavithana \cite{horawalavithana:graph}    & (2016)        & \makecell[c]{}                        &                       &     &                   &       \\

% Zhang \cite{}
% \cite{zhang}                                    &               &                                       &                       &     &                         &       \\
%
\bottomrule
\end{tabular}
\caption{Summary of the features of selected works sorted by publication date.
$^1$\textbf{Generic Design}: this criterion indicates whether a given scheme
provides a graph processing framework that supports more than one graph algorithm (\faThumbsOUp)
or whether it focuses on concrete graph algorithm(s) (\faThumbsDown). 
$^2$\textbf{Considered Problems}: this column lists graph problems (or algorithms)
that are explicitly considered in a given work; they are all explained in~\cref{sec:back_problems}.
$^3$\textbf{Used Programming Paradigm, Model, or Technique}: this column specifies programming paradigms and models used
in each work; they are all discussed in~\cref{sec:back_models} and~\cref{sec:fpga-techniques}. ``None'' indicates that a given
scheme does not use any particular general programming model or paradigm or technique.
$^4$\textbf{Multi FPGAs}: this criterion indicates whether a given scheme scales to
multiple FPGAs (\faThumbsOUp) or not (\faThumbsDown).
$^5$\textbf{Input Location}: this column indicates the location of the \emph{whole} input graph dataset.
``DRAM'', ``SRAM'', or ``HMC'' indicates that it is located in DRAM, SRAM, or HMC,
and it is streamed in and out of the FPGA during processing (i.e., only a part of
the input dataset is stored in BRAM at a time). Contrarily, ``BRAM''
indicates that the whole dataset is assumed to be located in BRAM.
``Hardwired'' indicates that the input dataset is encoded in the FPGA reconfigurable logic.
$n^\text{\textdagger}, m^\text{\textdagger}$: these two columns contain the numbers of
vertices and edges used in the largest graphs considered in respective works.
%
% $^6$\textbf{Achieved MTEPS}: MTEPS (Million traversed edges per second) is a value often used to compare
% existing traversal algorithms such as BFS; it is the number of graph edges traversed per second.
% We report the
% highest value that was measured on a single FPGA in each considered work.
%
In any of the columns, ``unsp.'' indicates that a given value is not specified.
} 
\label{tab:papers}
\vspace{-1em}
\end{table*}

\section{FPGA-Specific Graph Programming Techniques}
\label{sec:fpga-techniques}

We discuss separately graph programming techniques that are unique to FPGAs.

\subsection{Direct Hardware Mapping}
\label{sec:hardware_mapping}

One of the earliest approaches in FPGA graph processing was to to map a graph
completely onto the FPGA using logical units and wires to physically represent
every single vertex and edge.  The obvious limitation of all these approaches
is that the size of input graphs is limited by the amount of the FPGA
reconfigurable resources. 

In 1996, Babb et al.~\cite{babb1996solving} introduced \emph{dynamic computation
structures}, a technique that compiles a given graph problem to Verilog code
and then maps it to an FPGA. Each vertex of the input graph is physically
represented on the FPGA. This system was then able to solve the Bellman-Ford
algorithm to find all shortest paths from a specified source vertex. The
approach requires to reconfigure the FPGA whenever the input graph changes.

Later, Mencer and Huelsbergen~\cite{huelsbergen2000representation,
mencer2002hagar} presented a more flexible design where a graph topology can be
changed on the FPGA using tri-state buffers to represent entries of the
adjacency matrix.  This, however, still requires $O(n)$ space on the chip.  Simple
graph problems, such as reachability of some vertex $b$ from another vertex $a$,
can be solved by propagating a signal through $a$ and checking whether it
reaches $b$.  Other problems, such as SSSP or CC, can be solved in a similar
manner with slightly more complex circuits.

Finally, Dandalis et al.~\cite{dandalis1999domain} represent vertices as
processing elements but store edges in the main memory so that they can be
loaded dynamically to represent same-size graphs with different edge sets
without having to reconfigure the FPGA. Jagadeesh et
al.~\cite{jagadeesh2011field} implement a similar design but propose changes to
the design of the processing elements to reduce the number of cycles per
iteration. Both approaches can compute arbitrary graphs as long as the FPGA has
been configured with enough processing elements. 

\subsection{Graph Virtualization}

Graph Virtualization is a technique proposed by Lee et
al.~\cite{lee2017extrav}, where the program running on the host processor is
provided with the illusion that the input graph resides in the main memory and
is stored using some simple format such as Adjacency Array, while in reality
the graph is stored in a more complex, possibly multi-level and compressed form
on a storage managed by an accelerator such as FPGA.  The motivation is to use
the accelerator to offload tasks related with graph decompression and filtering
from the host processor. 
Additionally, graph virtualization enables the accelerator to apply various
optimizations and functionalities to the data, for example multi-versioning,
without affecting processor functions or programmability.
This technique can be used together with any accelerator, not only an FPGA.
However, as the proposed design is implemented on an FPGA system, we include it
in this survey.

\section{Specific Graph Algorithms on FPGA}
\label{sec:fpga-algs}

We now discuss selected hardware implementations of individual graph
algorithms. Such schemes form one significant part of research works dedicated
to graph processing on FPGAs.

\subsection{BFS}

Works on BFS constitute the largest fraction of graph
processing schemes on FPGAs. We now describe selected works, focusing on explaining key
ideas and summarizing important outcomes.

\subsubsection{Using Hybrid CPU-FPGA Processing}

\macb{One idea} for efficient BFS traversals is to combine the different
compute characteristics of the FPGA and the CPU.  Specifically, Umuroglu et
al.~\cite{umuroglu:hybrid_bfs_FPGA} present an efficient BFS implementation on
a CPU-FPGA hybrid system. The paper focuses especially on small-world
graphs~\cite{watts1998collective} which have a very low diameter. In such
graphs, the size of the frontier exhibits a specific pattern throughout the BFS
algorithm. The frontier remains fairly small in the first several iterations,
but then grows quickly in the next steps, to become small again at the end of
the traversal~\cite{beamer2013direction}. As the authors estimate, the frontier
contains on average about 75\% of all the vertices in the considered graphs
during the fourth iteration.
The authors use this observation while splitting the work between the CPU and
the FPGA: small frontiers do not offer much parallelism but can be efficiently
computed on the CPU while large frontiers can be parallelized on the FPGA. The
implementation described in the paper thus \emph{computes the first and last
steps on the CPU and uses the FPGA only to compute the steps with the large
frontiers}. The implemented BFS is expressed using the language of linear
algebra (i.e., frontier expansion is implemented as multiplying the adjacency
matrix and a vector that stores the current frontier,
see~\cref{sec:back-abstractions} for more details).

% The paper expresses BFS in terms of linear algebra operations which presents
% the problem in another view and shows how redundant computations can in fact
% lead to higher memory bandwidth.  Each BFS step is computed as a
% matrix-vector multiplication $x_{t+1}=A*x_t$, where $x_t$ is a bit vector,
% encoding the visited nodes at the $t$-th iteration and $A$ is the adjacency
% matrix.  The distance of a node is then set by comparing the $x_t$ and
% $x_{t+1}$ vectors: If a node has been visited in the $t$-th step for the
% first time, it has a distance $t$. 

\macb{Another idea} in the work by Umuroglu et al.~\cite{umuroglu:hybrid_bfs_FPGA} is
to \emph{read the whole frontier array sequentially}. In BFS, an important
operation is to verify whether a certain vertex is in the frontier.  Instead of
querying only for those vertices, the authors propose to read the whole
frontier array into BRAM and thus remove the need for random memory accesses.
Because of the small-world graph assumption, we know that the frontiers will
contain a significant amount of vertices. 

\macb{Various Insights}
The authors name three important ways for using a large portion of the
available DRAM bandwidth: A high rate of requests to make the latency of
individual requests less significant, using large bursts of requests, and a
sequential access pattern to increase the number of row buffer
hits~\cite{umuroglu:hybrid_bfs_FPGA}. The authors argue that it is more
efficient to treat sparse bit vectors as dense
(see~\cref{sec:back-sparse-dense} for an explanation of sparse and dense
structures) and read them sequentially instead of accessing (most often
randomly) only the parts that are know to contain the required data.

\macb{Remarks on Evaluation}
As the vector that is the result of the MV multiplication (i.e.,
the result of the frontier expansion) is stored in BRAM, the BRAM poses a hard
limit on the size of graphs that can be computed using this approach. In fact,
the authors were unable to use graphs with more than 2 million nodes and 65
million edges due to the limited on-chip BRAM size.  The size of the result
vector is $n$ words; the paper reports that 82\% of the BRAM is used for
the result vector. 

\subsubsection{Using Hybrid Memory Cubes (HMC)}

The \macb{key idea} due to Zhang et al.~\cite{zhang:graph_FPGA} is to use
Hybrid Memory Cubes (HMC) to implement an efficient BFS implementation on FPGAs
(we discuss HMC in more detail in~\cref{sec:back-HMC} and~\cref{sec:hmc}). 
The authors build an analytical performance model of the HMC memory access
latencies that allows them to analyze performance bottlenecks. HMC allows to
select the size of the transferred data payload to be anything between 16 bytes
and 128 bytes. The analysis shows that, depending on data locality~\cite{tate2014programming}, either
small or large read granularity is more efficient. For example, reading a list
of neighbors in a BFS traversal has better data locality than updating the
parent of a single vertex. Thus, \macb{the main insight} is that
\emph{different payload sizes for accessing different data structures in BFS
can be used to optimize performance}.

In contrast to Umuroglu et al.~\cite{umuroglu:hybrid_bfs_FPGA}, the authors do
not use a hybrid FPGA-CPU setting and run the whole BFS on the FPGA,
including iterations with smaller frontiers.  In such iterations, the bitmaps
are very sparse and entries should be accessed without having to iterate the
whole array. The \macb{proposed solution} is to use a level of indirection in a
form of a second smaller bitmap where each entry represents an OR reduction of
$k$~entries in the larger bitmap. This second bitmap is conveniently small
enough such that it can be stored in BRAM ($k$ can be set to be arbitrarily
large). 

\macb{Implementation Details}
The authors implement BFS in a framework based on the MapReduce paradigm.
{Mapper modules} read the current frontier of the BFS and extract the adjacency
lists of those vertices. {Reducer modules} then update the parents of the
newly visited vertices and store them in the new frontier. The frontiers are
stored as bitmaps in the main memory.

\macb{Remarks on Evaluation}
The described work enables processing arbitrarily large graphs. However, the 
used HMC module had a limited capacity of 4GB, limiting the size of used
graphs.

% The implementation was able to deliver up to 166 MTEPS. 

\subsubsection{Customizing Graph Representations}

The \macb{main idea} by Attia et al.~\cite{attia2014cygraph} is to \emph{use a
custom graph representation to reduce the memory bandwidth usage} in their BFS
design~\cite{attia2014cygraph}. The representation is based on the CSR
format. \macb{One observation} the authors make is that the information stored
in the $row$ array is only needed until a vertex has been visited for the first
time.  Thus, the authors propose to additionally store both the visited flag as
well as the distance of a visited vertex in the $row$ array. In their design,
the least significant bit of an entry $row[i]$ indicates whether the vertex~$i$
has been visited. If $i$ has been visited, the rest of the entry stores the
distance to the source vertex.  Otherwise, $row[i]$ stores the offset of its
neighborhood in $col$ as well as the number of neighbors.
In \macb{another optimization}, instead of pushing vertex to the frontier
queue~\cite{besta2017push}, the authors propose to push the value $row[i]$
since only this value is needed in the subsequent iteration. It should be noted
that this version of BFS only outputs the distance of each vertex from the
source but not their parents. All these optimizations improve spatial locality
and reduce the amount of data that has to be read from DRAM.

\subsubsection{Others}

Other implementations of BFS on FPGAs have been
proposed~\cite{wang2010message, ni2014parallel,
betkaoui2012reconfigurable}.
They come with various optimizations and schemes, for example Wang et
al.~\cite{wang2010message} deliver a message-passing multi-softcore
architecture with a variety of optimizations (dual-threaded processing, using
bitmaps for output, and pipelining), Ni et al.~\cite{ni2014parallel} combine
using SRAM chips and DRAM modules with optimized pipelining, and Betkaoui et
al.~\cite{betkaoui2012reconfigurable} decouple computation and communication
while maintaining a large number of memory requests in flight at any given
  time, to make use of the parallel memory subsystem. 
Finally, FPGAs are used to accelerate a distributed BFS traversal~\cite{bernaschi2012breadth}. Specifically, they implement virtual-to-physical address mapping
in a cluster of GPUs based on the Remote Direct Memory Access~\cite{gerstenberger2014enabling} technology.

\subsection{SSSP}

Tommiska and Skyttä~\cite{tommiska2001dijkstra} implement Dijkstra's SSSP
algorithm on the FPGA using VHDL. The input graph, represented as an adjacency
matrix, and the result data are stored on the internal storage of the FPGA,
which drastically limits the size of graphs that can be computed. \macb{An
interesting feature} of this design, which is also found in some later
works~\cite{zhou2015sssp}, is \emph{using the comparator block which is able to
process multiple edges in parallel to find the one with the smallest distance}. 

The \macb{key idea} in the work by Sridharan et
al.~\cite{sridharan2009hardware} is to use a linear-programming based solution
for certain graph problems on the FPGA, \emph{which forms the only work on
  solving graph problems on FPGAs using linear programming}.  However, their
  design also limits the size of graphs that can be processed; the authors
  could only test their implementation on a graph with 44 vertices. 

% Lam et al.~\cite{lam2010accelerating} describe a possible implementation of a
% data structure called bucket-heap in hardware which accelerates Dijkstra's SSSP
% algorithm and reduces the number of memory accesses. This work has not been
% implemented an FPGA yet.\maciej{check} 

Zhou et al.~\cite{zhou2015sssp} developed an FPGA graph accelerator for solving
SSSP using the Bellman-Ford algorithm.  The first \macb{key feature} of this
architecture is, unlike in previous approaches, \emph{storing the input graph
in DRAM}. The \macb{second key feature} is that the architecture is \emph{fully
pipelined} with the ability to process multiple edges concurrently; $p$ edges
are read in parallel and the corresponding data is driven through multiple
architecture blocks, each of which can process $p$ instances at once. 
The first stage is the sorting block, which makes sure that if more than one of
the $p$ edges target the same destination vertex, only the one with the minimal
value is used to produce an update. Next, the memory read block fetches the
weights of the targeted destination vertices from DRAM. These values are then
passed to the computation block, which compares the proposed updated weight
(i.e, the new distance) and the current weight of the destination vertices and
decides whether to perform the update. Finally, the memory write block stores
the updated values to DRAM. It also counts the number of successful updates
during one whole iteration and makes sure to terminate the algorithm if no more
updates have occurred. 
Finally, the \macb{third key part} of the architecture is a \emph{data-forwarding
technique} to prevent race conditions that can occur when two edges with the
same destination are processed in consecutive cycles. The authors propose to
sort the edges by their destination vertices since all updates to the same
vertex would follow consecutively and the minimal value could be found easily
in the sorting block, thus reducing the number of updates that have to be
written to DRAM.  The performance limitation in the design is the bandwidth and
the high number of random memory accesses that have to be made. 

A slightly different approach is taken by Lei at al.~\cite{lei2016fpga}, using
a version of the so called ``eager'' Dijkstra
algorithm~\cite{edmonds2006single} instead of Bellman-Ford. The algorithm works
with a priority queue and can be parallelized by removing multiple vertices
from the queue in each iteration. A similar priority queue such as in work by
Sun and Srikanthan~\cite{sun2006accelerating}, called systolic array priority
queue (SAPQ), is implemented on the FPGA. However, to support the processing of
large graphs, elements that overflow the queue are moved to DRAM.  The benefit
of this queue implementation is that enqueuing an element and dequeuing the
minimum element are performed in constant time.  
In each iteration, $\lambda$ vertices are picked from the priority queue and
processed in one of $\lambda$ processing elements in parallel.  Similarly to
Zhou et al's approach~\cite{zhou2015sssp}, this design suffers from the large
amount of off-chip memory accesses. The authors derive a performance model
based on the total number of memory accesses. Namely, in each iteration, $(512
+ 64 d)\lambda$ memory accesses are required for reading the graph data and
$1024 d \lambda$ memory accesses are needed to store the results ($d$ is the
average degree in the input graph). It should be noted that one iteration in
this case corresponds to reading $\lambda$ vertices from the priority queue and
\emph{not} traversing the whole data structure. Since different DRAM modules
are used to store the graph data, the priority queue, and the results, the
performance bottleneck depends on the module which is used most.  Thus,
assuming that the bandwidth is the limiting factor, the total processing time
is $1024 d L \lambda \backslash B_{DRAM}$ for $L$ iterations. 

\subsection{APSP}

Bondhugula et al.~\cite{bondhugula:APSP_FPGA, bondhugula2006hardware} solve the
APSP using a \emph{parallel version of the Floyd-Warshall algorithm}. The input
graph is stored in off-chip memory and only the required \emph{slices}
(i.e., parts of the adjacency matrix) are streamed to the FPGA BRAM modules.
The tiling scheme by Venkataraman, et al.~\cite{venkataraman2003blocked} is
used to partition the graph into multiple tiles of size $B \times B$, where $B$
is a scheme parameter. The architecture contains $B$~pipelined processing
elements, with each one being able to process up to $l$~values from a row per
cycle.  These values are passed from one processing element to another until
they are finally stored back to the memory by the last processing element. The
tile size is limited by the resources available on the FPGA. 

Betkaoui et al.~\cite{betkaoui:APSP_FPGA} solve APSP for unweighted
graphs by \emph{running BFS from each vertex}. The design
contains multiple processing units, each of which is able to run one BFS
instance in parallel with others. Each processing unit has its own local memory
but is also connected to a shared memory system where the graph representation
is stored.  The \macb{key idea} (for optimizing memory accesses) is to
\emph{use the on-chip memory for data requiring irregular patterns} (e.g., the
current status of vertices) and \emph{the off-chip memory for data that can be
accessed sequentially} (e.g., the edges of the graph). For this, they design a
bitmap scheme that can efficiently store the visit status of vertices (i.e.,
whether each vertex was visited) and emulates queues used in the algorithm.
Since each of the three bitmaps contain at most $n$~bits, the authors make the
assumption that these bitmaps can fit in the on-chip memory. However, since
each of the $p$ processing element stores three bitmaps of size $n$, the total
space requirement is $3pn$. Thus, this assumption does not hold for larger
graphs. 

% The approach by Betkaoui et al.  is more suitable for sparse graphs while the
% solution by Bondhugula et al.  is better suited for dense data sets.
% \maciej{verify}

\subsection{PageRank}

The only work which describes an FPGA accelerator specifically tailored for
PageRank was done by Zhou et al.~\cite{zhou2015pagerank}.  The authors use a
two-phase algorithm (based on the GAS model, see~\cref{sec:back_models})
similar to their general approaches in FPGA graph processing
frameworks~\cite{zhou2016high, zhou2018framework}, which we discuss in more
detail in~\cref{sec:zhou_prasanna_framework}.  During each iteration, the
scatter phase processes all edges, generates updates, and stores the updates in
off-chip memory. The gather phase reduces the updates and applies them to
the array with ranks. The \macb{main focus} of the approach is to reduce the
number of random memory accesses during the algorithm. This is achieved by
reading the edges and the vertex properties in the scatter phase sequentially,
reading the updates and writing the updated PageRank values in the {gather
phase} sequentially, and reducing the number of random memory writes in the
{scatter phase} by leveraging the order in which the edges are being processed
to merge some of the updates before writing them to DRAM. 

\subsection{Application-Driven Graph Problems}

Finally, there have been attempts at implementing various
specific graph-related applications on FPGAs. An example is
DNA Assembly based on De Bruijn graphs~\cite{compeau2011apply}
using FPGAs~\cite{varma2013fassem, poirier2018dna}.

\section{Generic Graph Processing Frameworks on FPGA}
\label{sec:fpga-frameworks}

A lot of effort was placed in designing generic frameworks for facilitating
the implementation of graph algorithms on the FPGA. The goal of such a
framework is to (1) allow the user to easily and rapidly develop their graph
algorithms, using some proposed graph programming abstraction or programming
model, and (2) generate an FPGA implementation from such user code.  This
yields \macb{two major advantages} to custom FPGA designs: \emph{portability} and
\emph{programmability}. First, the user algorithm is not tied to one specific
FPGA device but can be compiled to various FPGA devices. Second, the framework
is usually built around a specific programming model and any graph algorithm
that is compatible with this model can be implemented without having to change
the core of the framework.  
In the following, we present selected FPGA graph frameworks chronologically.

\subsection{[2006] GraphStep \cite{kapre2006graphstep}}

GraphStep by Kapre et al.~\cite{kapre2006graphstep} is \emph{one of the first
approaches to design a generic FPGA architecture} that is not specific to only
one graph algorithm. In GraphStep, every vertex is abstracted to be an
\emph{actor} that can send messages and invoke selected pre-programmed methods
on its neighboring vertices. The computation associated with each vertex is
divided into several steps; these steps are synchronized across all the
vertices. First, each vertex receives input messages coming from its neighbors.
Second, a vertex awaits the synchronization barrier. Third, a vertex updates
its state. Finally, a vertex sends messages to its neighbors. 

%\macb{Programming Model}
%
The used model resembles BSP, with a difference that communicating objects
are graph vertices and communication can only occur between adjacent
vertices. It also resembles the vertex-centric model introduced later in
the Pregel paper~\cite{malewicz2010pregel}.  Such a model can be realized in
several ways, among others, being hardware mapped like the early works of Bobb
and Mencer (see~\cref{sec:hardware_mapping}) where every vertex is physically
represented on the FPGA, or by having multiple processing elements on the FPGA,
with each element taking care of a certain range of vertices.

%\macb{Optimizations}
%
High-degree vertices are decomposed into multiple
vertices with smaller degrees to avoid bottlenecks. Additionally, a selected
graph partitioning scheme that minimizes sizes of
cuts~\cite{caldwell2000improved} is used to maximize the locality of vertices
stored on the same memory block. 

%\macb{Hardware Design}
%
Multiple processing
engines are distributed on the FPGA, each of them being able to process one
edge per cycle using a pipelined architecture. One such processing engine
needs 320 Virtex-2 slices, and thus a XC2V6000 FPGA can hold around 32 such
processing engines, connected with a Butterfly
topology~\cite{dally2004principles, besta2014slim}. 
The input graph data resides in BRAM instead of
external memory since the FPGA on-chip bandwidth is one to two orders of
magnitude higher than the off-chip bandwidth to and from DRAM. For larger
graphs that do not fit the on-chip memory, they propose to distribute the data
among multiple FPGAs. This way the bandwidth would scale with the size of the
data.  
64K edges can be stored on the BRAM of one XC2V6000 FPGA and hence 16 such
FPGAs are required to store and process the whole ConceptNet. 48 additional
FPGAs are used to route the communication. In total, 64 FPGA are required to
solve a graph problem with 550,000 edges; only 25\% of the incorporated FPGAs
perform the actual computation.

%\macb{Example Use Cases}
%
GraphStep implements the spreading activation of the ConceptNet knowledge
base~\cite{liu2004conceptnet}, as an example application of their framework. 

%\macb{Remarks on Evaluation}
%
The authors compare their approach with a sequential 
implementation on a 3.4GHz Pentium-4 Xeon and obtain speedups 
of 10--20$\times$.
%
%depending on the size of the query into the graph. 
% 
% Queries with more words 
% lead to a superior speedup compared to the CPU implementation.  
% 
% Computing a graph with half a billion edges would require 
% at least 1000 times more Virtex-2 FPGA devices, of which most are only used 
% to route the communication, but don't do any real computation. 
% Thus the approach by GraphStep to store all graph data on the FPGAs seems 
% largely infeasible when considering larger graph data sets. 

\subsection{[2011] Framework by Betkaoui et al. \cite{betkaoui2011framework}}

The next FPGA graph processing framework is built upon a vertex-centric
approach. The authors use the graphlet counting problem as a case study.  The
leading design choice behind their work is to maximize the utilization of the
available parallelism by implementing a large number of processing units on the
FPGA. All processing units have a direct access to the shared off-chip memory,
which consists of multiple memory banks, to increase throughput and allow
memory operations to be performed in parallel. This approach can even be
scaled to multiple FPGA units. 

On the other hand, no form of caching is used and the data is not prefetched to
BRAM. This leads to large latencies when accessing data from DRAM. These
latencies are alleviated by and their cost is outweighed with the large amount
of used parallelism.

The authors implement their design on the Convey HC-1, a hybrid system,
featuring an Intel Xeon dual-core host CPU and four programmable Xilinx Virtex
5 FPGAs~\cite{austin2010convey}. The FPGAs are connected to 16 DDR2 memory
channels over a crossbar. There are 8 independent memory controllers with
connections to each device, each one implemented on a V5LX110 FPGA. The memory
itself consists of 16 special Scatter-Gather-DIMMs, totaling 8GB of memory and
a peak bandwidth of 80 GB/s. These memory units offer very good random memory
access performance. 

%Interestingly, the efficiency scales almost
%linearly when increasing the number of processing units. 
%
%By using 128 instead
%of 64 processing units, the efficiency drops to 90\%. 
%
%In contrast, CPU
%implementations scale much worse when increasing the number of cores. 

\subsection{[2014] GraphGen \cite{weisz:GraphGen}} 

GraphGen is a framework that can compile graph algorithms, which are expressed
according to the vertex-centric programming paradigm, onto a target FPGA. The
main focus is on enabling developers to program graph applications on hardware
accelerators without requiring any knowledge about the specific platform. 

In contrast to previous graph frameworks, GraphGen allows to store graph data
in DRAM, thus enabling the processing of much larger graphs that would
otherwise not fit in BRAM. For this, the graph is partitioned (according to the
1D graph partitioning scheme~\cite{boman2013scalable}) into smaller subgraphs, each of which
can fit in BRAM.  The compiler generates an FPGA program for each subgraph. The
compilation procedure is thus dependent on both the graph algorithm and the
graph data, which means that the FPGA kernel needs to be recompiled whenever the
graph data changes or the user wishes to compute the same algorithm on another
data set. 

The computation revolves around an update function that is executed on every
vertex of the graph. This function is specified as a series of custom graph
instructions.  The user needs to define these instructions and provide
appropriate RTL implementations for them. They are then integrated in the graph
processor and used by the compiler to construct the update function. 

The architecture is based on the CoRAM abstraction~\cite{chung2011coram},
a memory abstraction model for FPGAs. The model specifies buffers which are used 
for the transfer of data between the FPGA and other devices. The data transport 
between CoRAM buffers and external DRAM is then specified using a C-like language 
and managed by so called control threads.

The framework applies several additional optimizations to reduce the bandwidth
usage.  First, double buffering is used to prevent stalls while waiting for new
data.  Second, edges are sorted according to the order in which they are
accessed. Thus, memory access requests can be coalesced to reduce the number of
independent memory accesses.  Third, pipelining is used in the processing
elements to achieve high clock rates. The GraphGen compiler schedules
instruction such that data hazards are avoided.  Moreover, the incorporated
programming model allows multiple edges to be read the same cycle.  Finally,
custom SIMD (single-instruction multiple-data) instructions are supported to
allow processing data elements together. 

The authors provide two case studies for GraphGen, stereo matching using the
Tree-Reweighted Message Passing algorithm (TRW-S)~\cite{sun2003stereo} and
handwriting recognition using a Convolutional Neural Network
(CNN)~\cite{o2016neural, distdl-preprint, deep500}. When compared to software implementations, they
achieve a speedup of 14.6$\times$ and 2.9$\times$ in FPGA implementations
compiled by GraphGen for TRW-S and CNN, respectively.  Both applications come
from the computer vision domain and work on 2D images that are modeled as 2D
grids.  The advantage of such graph data sets is their regularity: every pixel
is only adjacent to the eight other pixels around it. Thus, it is relatively
straightforward to split the graph into subgraphs while still maintaining high
locality.

\subsection{[2015] GraphSoC~\cite{kapre:SoCs_FPGA, kapre:custom_graph_FPGA}}

GraphSoc~\cite{kapre:SoCs_FPGA, kapre:custom_graph_FPGA} is an FPGA graph
processing framework that comes with an instruction set architecture for
performing algorithms on sparse graphs. The input graph is stored entirely on
the FPGA BRAMs using the CSR format.
The FPGA contains multiple processing elements which are responsible for
processing subsets of the graph and are connected with a Network-on-Chip (NOC)~\cite{besta2018slim}.
Each processing element is implemented as a 3-stage pipeline in which an
instruction is fetched, decoded and then executed. Instead of having
general-purpose registers, the architecture has special dedicated registers for
vertices, edges, and instructions. There are also registers dedicated to
counting loops that allow zero-overhead looping.

Graph algorithms are expressed using the BSP model. There are four customizable
instructions, \emph{send, receive, accum}, and \emph{update}, that can be
specified to implement the needed functionality of a wide range of algorithms.
%
%They are implemented as $\mu$coded datapaths which are generated from C++ code
%using a custom assembler. 

The authors implement GraphSoc in high-level C++ and use High-Level Synthesis
(HLS) to generate RTL. They also use the PaToH
partitioner~\cite{catalyurek1999hypergraph} to minimize the number of
inter-partition connections when distributing the vertices among the processing
elements. 

If the data set is too large to fit in the BRAM, multiple FPGA devices
are used. For this case, the authors describe the design of a Beowulf cluster
of 32 Zync Z7020 boards for the computation of large sparse
graphs~\cite{kapre:SoCs_FPGA}. 
% 
% We do not provide more details of this design
% because the computation is partly done on the embedded ARM processors and is
% thus outside the scope of this survey. 

\subsection{[2016] FPGP \cite{dai:fpgp}}

FPGP~\cite{dai:fpgp} is a framework that uses \emph{intervals} and
\emph{shards} to improve locality of processed graphs and to enable the
processing of arbitrarily large data sets by storing only the currently
relevant graph partitions in BRAM.  Vertices are partitioned into $P$ intervals
based on their vertex IDs. Each interval has a corresponding shard (also called
a \emph{bucket}) that contains all edges that point towards one of the vertices
in this interval.  A shard is further divided into sub-shards, based on the
source vertex of the edges. Edges inside each sub-shard are then sorted by
their destination vertex. 
This partitioning scheme is chosen so that the order of updating vertices
corresponds to the intervals. This makes it possible to have multiple
processing units on an FPGA or even multiple FPGAs compute different intervals
in parallel.  While the edges are streamed from memory shard by shard, the
vertices are computed interval by interval. 
%
% first all vertices in the 
% first interval are updated, then all vertices in the second internal, and so on. 
%
This partitioning scheme is further described in a paper describing the NXgraph
processing system, developed by the same authors~\cite{chi2016nxgraph} (NXgraph
is a graph processing system designed for efficient graph computation on single
servers).

The FPGP framework enables the user to express different graph algorithms by
only implementing the \emph{kernel function}, which is executed for each edge
by dedicated processing units. The kernel function takes as input the
properties of the source and destination vertices of an edge and computes the
new property of the destination vertex. During the computation of an interval
the updated vertices are stored locally on in BRAM and then later written to
the shared memory together with other vertices in the interval. 

The general scheme behind the framework works as follows. For every
interval~$I_j$, the framework processes all edges in the corresponding
consecutively stored sub-shards $S_{i,j}$, where $i$ denotes the interval to
which the sources vertices of the edges belong. For every sub-shard~$S_{i,j}$,
the properties of the vertices in interval~$i$ are loaded to the FPGA and the
kernel function is performed for every edge in the sub-shard.  After the whole
shard is processed, the updated interval~$I_j$ is stored back to the shared
memory and the next interval is loaded to BRAM. Depending on the algorithm,
different numbers of iterations are needed.  Thus, the entire source property
array is being loaded $P$ times. Since $P$ can be expressed as
$\frac{n}{M_{BRAM}}$, $M_{BRAM} $ being the available BRAM memory on the chip,
the total communication cost for reading vertex properties is in $O(n^2)$. 

The authors build a model to describe the performance of the
framework.  They derive formulas that describe the effect of several factors on
the performance, including limited bandwidth. FPGP uses two different memory
systems: local DRAM banks that store the edges and a shared off-chip DRAM that
stores the vertex properties.  The communication for loading the vertex
properties used for one iteration of an algorithm is given as
$\frac{P n b_v}{N_{chip}B_{share}}$ where $b_v$ is the number of bits per vertex
property, $N_{chip}$ is the number of FPGAs, and $B_{share}$ is the bandwidth
to the shared memory.

% \begin{itemize}
%   \item The communication cost rises quadratically with $n$ which leads to bandwidth congestion for larger graphs. 
%   \item The communication cost can be reduced up to a factor of $P$ by using devices with more BRAM to store larger 
%         intervals, and in the best case the whole graph, on the chip. 
%   \item To avoid a quadratic slowdown, the BRAM memory of a device should grow with the size of the data set. 
% \end{itemize}

The authors evaluate the performance of FPGP with a BFS implementation on both
the Twitter graph and the Yahoo web graph and compare the results with
frameworks built for the CPU to show that their FPGA implementation can
outperform CPU implementations. 

% calculate the performance to be $frac{m}{fP^2}$, 
% where $f$ is the frequency of the FPGA. H

\subsection{[2016] GraphOps \cite{oguntebi:GraphOps}}

In GraphOps, the authors provide a
set of building blocks from which graph algorithms can be composed.  These
blocks are then placed on the FPGA and connected together. Example blocks
include {ForAllPropRdr} (it fetches the vertex properties of all the
neighbors of a vertex from memory to the FPGA), {NbrPropRed} (it
performs a reduction on the vertex properties of a neighborhood. The reduction
can be customized depending on the algorithm), and {ElemUpdate} (it
writes the updated value of a vertex back to memory).
The authors show how to use these three blocks and some utility and
control-flow blocks to implement PageRank and other algorithms. 

For example, during one iteration, the PageRank algorithm would execute the
three operations for every vertex in the graph. The first block reads the
vertex properties from memory, then this data is passed to the second block
to perform the reduction. Finally, the last block receives the updated value
and the address of the vertex and updates the memory location accordingly. 

The authors implement a range of different graph algorithms to illustrate the
flexibility of GraphOps. In their evaluation, they conclude that the bottleneck
for all implementations is the memory bandwidth between FPGA and DRAM.
  Parallelism is only available in the pipelined architecture of the blocks and
  inside individual blocks.

GraphOps does not describe how to implement multiple parallel pipelines.
Moreover, data that is fetched from memory is only used to update a single
vertex, but never reused among multiple vertices.  Such reuse could potentially
lower the communication costs and thus improve the performance. Since multiple
blocks can issue memory read or write requests simultaneously, the memory
channel needs to serve different memory interfaces. This, as well as the fact
that the neighborhoods are not accessed in the order in which they are stored
in memory, leads to a large number of random memory requests which place an
even harder burden on the memory bandwidth.  The authors conclude that
performance of GraphOps corresponds to only $1/6$ of the theoretically
available throughput.

To alleviate the memory bandwidth bottleneck between FPGA and DRAM, GraphOps
uses a locality-optimized graph representation which trades redundancy for
better locality and thus overall less communication (see
also~\cref{sec:locality_optimized}).

Instead of storing the properties of vertices as an array of size $n$, they
propose to replicate the property value for every single incoming edge.  The
``\textit{locally optimized property array}'' is thus an array of size~$m$
where every entry~$i$ represents the property of the destination vertex of the
edge that is stored in entry~$i$ of the adjacency array.  This design has two
major benefits. First, it removes one level of indirection when accessing the
property of an adjacent vertex. In a traditional adjacency array, one first
needs to access the index of the adjacent neighbor and then use it to access
the property array. In GraphOps, one can directly access the property of a
neighbor without having to know its ID.  Second, the properties of all
neighboring vertices are now stored consecutively in memory. This improves
spatial locality as there is now only one random memory access needed to access
all properties of adjacent vertices.  On the other hand, updates have to be
propagated to the replicated entries.

\subsection{[2016] GraVF \cite{engelhardt2016gravf}}

GraVF~\cite{engelhardt2016gravf} offers flexibility and ease of programming by
only specifying two functions used as building blocks for graph algorithms.
GraVF is based on the BSP model. In each superstep, the {Apply}
function defines how a vertex property is updated based on the incoming
messages and the {Scatter} function defines which messages are sent to
the vertex neighbors. Vertices are divided among the processing elements on
the FPGA. The synchronization step is implemented with a barrier algorithm
similar to that of Wang et al.~\cite{wang2010message}.  All vertices and edges
are stored on the FPGA, limiting the size of graphs that can be processed by
the framework. The largest graph that the authors were able to test on a Xilinx
Virtex 7 had 128k vertices and 512k edges. 

\subsection{[2017] ForeGraph \cite{dai:foregraph}}

In their second FPGA graph framework, ForeGraph~\cite{dai:foregraph}, Dai and
others focus on efficient scaling to multiple FPGAs. The driving observation in
their work is that multiple FPGAs enable more on-chip storage, more
parallelism, and more higher bandwidth.
The solution behind ForeGraph is to use a separate off-chip DRAM module for
each FPGA.  Similarly to the strategy in FPGP, the input graph is partitioned
into $P$ intervals and shards. However, now every partition corresponds to one
of $P$ FPGA devices. Each FPGA is thus processing the edges that lead to its
share of the graph but needs to access the source vertex properties from other
FPGA devices.  Since the vertex properties of a whole partition may not fit in
the BRAM of its FPGA, the partitions are further divided into $Q$ sub-blocks in
the same way. 

Both FPGP and ForeGraph are based on the assumption that the updated vertex
properties can be computed in intervals and that after all
edges leading to some interval~$i$ have been processed, the vertex properties
of the interval would not have to be updated again until the next iteration of
the algorithm. However, there exist graph algorithms that make an update of
some vertex~$w$ when traversing an edge $(u, v)$ where neither $u=w$ nor $v=w$
and also $w$ does not lie in the same interval as $v$.  An example is the
Shiloach-Vishkin algorithm where previous components have to be accessed and
reassigned~\cite{shiloach1980log}.  This and similar cases are handled by a
graph processing paradigm where all updates are first collected in some
temporary memory and then merged together. An example would be edge-centric
frameworks from Zhou and Prasanna~\cite{zhou2018framework}. 

% Not fpga related, so I only moved the idea about using the power law 
% distribution to the section about optimizations. 
% \subsection{TuNao \cite{j_zhou:TuNao}}
% TuNao is graph accelerator framework that is not directly related to FPGAs. 
% model. The architecture contains three modules, responsible for the three 
% stages, \textit{gather}, \textit{apply} and \textit{scatter}. Each module 
% is built with an Elastic Coarse-Grained Reconfigurable Array (ECGRA) 
% \cite{huang2013elastic}, which 
% To decrease the bandwidth usage the authors propose to store vertices 
% with very large degrees in BRAM. Intuitively these vertices 
% will be accessed much more frequently, thus by storing them in BRAM one can 
% exploit temporal locality. This follows from the skewed power-law distribution 
% which is often observed in real-world graphs \cite{faloutsos1999power}. 

% \subsection{[2017] ExtraV \cite{sth}}

\subsection{[2016 -- 2018] Frameworks by Zhou and Prasanna \cite{zhou2016high, zhou2017accelerating, zhou2018framework}}
\label{sec:zhou_prasanna_framework}

Throughout multiple papers Zhou and Prasanna follow the edge-centric paradigm
where, instead of accessing neighborhoods of vertices, all edges are streamed
to the FPGA and processed in the order in which they arrive, similarly to the
well-known system X-Stream~\cite{roy2013x}.  This results in sequential reads
to the graph data structure and leads to better bandwidth usage. As the edges
are sorted by their source vertex, the required vertex properties can also be
read sequentially by buffering small intervals of vertex properties in BRAM. 

The framework follows the Gather-Scatter programming model. In the first
(scatter) phase, all edges are processed and the resulting updates are stored
into buckets, sorted by the interval in which the updated vertex is located.
Because of the limited size of BRAM modules, these buckets reside in DRAM and
thus random memory writes are required.  In the second (gather) phase, the
updates are again streamed to the FPGA, merged together using an
algorithm-specific reduction function and then written to the vertex property
array in DRAM. 

% The edge-centric framework by Zhou et al.\cite{zhou2018framework} is built around the 
% scatter-gather-apply model. In the scatter phase every traversed edge is processed 
% and produces an update which is then applied in gather phase. Thus it is possible 
% to implement custom graph algorithms by simply programming the two functions 
% \textit{Process\_edge()} and \textit{Apply\_update()}. 

Random writes to DRAM often lead to row-conflicts, taking place when two
consecutive memory accesses target different rows in
DRAM~\cite{jacob2010memory}. To reduce the number of row conflicts in the
scatter phase, the authors propose to sort the edges inside each partition by
their destination vertex. Thus, updates to the same row in DRAM happen
consecutively. Given that there are $P$ partitions, one can show that there
will be at most $O(P^2)$ row-conflicts.  Additionally, since now updates that
target the same destination vertex are processed consecutively, these updates
can be reduced on the FPGA before being stored to the update bucket, reducing
communication. 

To lower power consumption, the authors temporarily deactivate BRAM modules
through the 'enable' port~\cite{BRAM_power_consumption} when they are not
needed.  Parallelism is implemented in the form of concurrent pipelines.  Each
pipeline can fetch and process a new edge in each cycle. There are three stages
in each pipeline which perform slightly different for the scatter and the
gather phase. In the scatter phase, the pipeline first reads the property of
the source vertex from BRAM, then computes the update and finally writes the
update to DRAM. In the gather phase, each pipeline reads an update message,
reads the destination vertex and its property, performs the reduction, and
stores the property back to BRAM. After all updates of an interval have been
processed, the vertex properties are written back to DRAM. 

One problem with the approach occurs when multiple pipelines concurrently 
process updates with the same destination vertex. For this case, the authors 
implement a combining network which reduces concurrently processed updates 
in case they have the same destination.   

The authors observe that some algorithms, such as BFS or SSSP, are not a
perfect fit for the edge-centric model as during each iteration of the
algorithm only certain parts of the graph needs to be accessed. Thus, in an
edge-centric model, several iterations through the edge list would need to take
place (e.g., $D$ iterations in BFS), and in many of these iterations only a few
edges would actually be needed.  Consequently, similarly to the observation by
Umuroglu et al.~\cite{umuroglu:hybrid_bfs_FPGA}, the authors take into account
that the size of the frontier can vary drastically between different
iterations. They observe that if the frontier is large enough, the edge-centric
iteration is still more performant than randomly accessing only the required
neighborhoods of the vertices in the frontier. However, for smaller frontiers,
the vertex-centric approach is more efficient.  Thus, they implement a hybrid
approach that dynamically switches between the two paradigms based on the
number of vertices in the frontier~\cite{zhou2017accelerating}. 

\section{Key Problems and Solutions}

In this section, we separately present the key problems that make graph
processing on FPGAs challenging and we discuss how they are approached by
various works. We focus on two key problems that are addressed by the vast
majority of works dedicated to graph processing on FPGAs: insufficient
bandwidth between the FPGA and DRAM (\cref{sec:problem_bw}) and low internal
FPGA storage (\cref{sec:problem_low-internal}).  Both of these challenges are
strongly related in that most schemes that address one of them are also
suitable for the other one. Thus, the structure of this section is not rigid.
%
%In particular, there are two complementary problems, where solving one makes
%the other irrelevant. 

\subsection{Insufficient DRAM--FPGA Bandwidth}
\label{sec:problem_bw}

The first key problem is the low bandwidth between the FPGA and the main
memory. This which makes graph processing memory-bound as most graph algorithms
perform relatively simple computations on large amounts of data. Moreover, many
graph algorithms require significant numbers of random accesses to
DRAM~\cite{besta2017push}. This incurs further overheads as randomly accessing values
from DRAM is substantially slower than loading data that is stored
consecutively. 
% 
% Moreover, accessing a value requrires multiple random memory accesses
% depending on each other. For example for accessing the vertex property of the
% $i$th neighbour of some vertex, one needs to first query the offset array,
% then the adjacency array and finally the vertex property array. Each of the
% three memory requests can only be started after the previous one was served. 
%
We now present several solutions proposed in the FPGA literature for this
particular issue.

\subsubsection{High-Performance Memory Interfaces}
\label{sec:hmc}

Several works use modern memory interfaces that enable high bandwidth. The
Hybrid Memory Cube (HMC) technology dramatically improves the bandwidth of
DRAM.  Benefits for graph processing on FPGA coming from HMCs are due to (1)
the HMC substantial bandwidth improvement over traditional DRAM, (2) optimized
parallel random memory access, and (3) in-memory locking and atomic operations.
Now, HMCs are used by various FPGA solutions for graph processing, starting
from the work by Zhang et al.~\cite{zhang:graph_FPGA}.  Two more recent
approaches have been able to further increase the performance of graph
algorithms on FPGAs equipped with HMC by using several techniques such as
relabeling indices according to their degrees~\cite{zhang2018degree} and
incorporating degree-aware adjacency list
compression~\cite{khoram2018accelerating}.

% Another similar technology is High Bandwidth Memory (HBM), which is
% implemented on AMD and NVIDIA GPUs and used for graph processing.

\subsubsection{Near-Data Computing on FPGA}

Certain efforts propose to implement near-data computing architectures on FPGAs
to ensure area and energy efficiency. For example, Heterogeneous Reconfigurable
Logic (HRL)~\cite{gao2016hrl} is a reconfigurable architecture that implements
near-data processing functionalities. HRL combines coarse-grained and
fine-grained reconfigurable logic blocks.  Among its other features are the
separation of routing network into ones related to data and control signals.
The former is based on a static bus-based routing fabric while the latter uses
fine-grained bit-level design.  Moreover, HRL uses specialized units for more
efficient handling of branch operations and irregular data layouts.
Now, the authors evaluate HRL on a wide spectrum of workloads, among others
graph algorithms such as SSSP.  HRL improves performance (per Watt) by
2.2$\times$ over FPGA and 1.7$\times$ over coarse-grained reconfigurable logic
(CGRA), and it achieves 92\% of the peak performance of a native near-data
computing system based on custom accelerators for graph analytics and other
workloads.

\subsubsection{Data Redundancy for Sequential Accesses}
\label{sec:locality_optimized}

As loading consecutive memory words is significantly more efficient that random
memory accesses, some FPGA works focus on developing graph data layouts that
exhibit more potential for sequential accesses. One example is the
GraphOps~\cite{oguntebi:GraphOps} graph processing framework, where instead of
storing the properties of vertices as an array of size~$n$, the property values
are replicated for every single incoming edge.  Other examples are graph
processing frameworks and implementations of specific algorithms where edges
are streamed between the FPGA and the main memory, according to the
edge-centric graph programming model~\cite{zhou2016high, zhou2017accelerating,
zhou2018framework}.

% The ``\textit{locally optimized property array}'' is thus an array of size
% $m$ where every entry $i$ represents the property of the destination vertex
% of the edge that is stored in entry $i$ of the adjacency array.  This design
% has two major benefits. First, it removes one level of indirection when
% accessing the property of an adjacent vertex. In a traditional adjacency
% array one first needs to access the index of the adjacent neighbor and then
% use it to access the property array. Here we can directly access the property
% of a neighbor without having to know its id.  Second, the properties of all
% neighboring vertices are now stored consecutively in memory. This improves
% spatial locality as there is now only one random memory access needed to
% access all properties of adjacent vertices.  Updates on the other hand have
% to be written to the normal property array and after every iteration of the
% algorithm, these updates have to be propagated to the replicated entries. 

\subsubsection{Merging Updates on FPGA}

A common step in graph algorithms is the reduction (i.e., merging) of updates
that target the same vertex. For example, in BFS, multiple vertices in the
frontier might be connected with the same vertex, but only one of them should
be declared as its parent. Another example is PageRank, where several updates
to a vertex are summed. Writing all these updates to DRAM unnecessarily
stresses the available bandwidth as the updates could already be reduced on the
FPGA before being stored to DRAM.

This solution to the limited bandwidth problem is proposed by Zhou et
al.~\cite{zhou2018framework} in their optimized data layout that allows to
combine some updates directly on the FPGA such that only one value has to be
written to DRAM. In their edge-centric framework, the graph is split into
partitions according to the source vertices (i.e., vertices with identical
source vertices are placed inside the same partition).  Then, the edges inside
each partition are sorted by their destination vertex. Thus, \emph{edges with
the same destination vertex are processed consecutively and the corresponding
updates be combined}. More details on this framework is provided
in~\cref{sec:zhou_prasanna_framework}.

% Without such a data layout, one would need complex data structures to group
% together the updates that target the same vertex.  This approach is also
% useful for avoiding conflicts when processing a partition in parallel. The
% edges of the partition can be distributed among processing units working in
% parallel in such a way, that edges with the same destination vertex are
% handled by the same processing unit. 

Dai et al.~\cite{dai:fpgp, dai:foregraph} reduce updates that target the same
graph partition on the FPGA BRAM.
% 
% until all relevant updates have been produced and only then write the reduced
% updates back to BRAM. 
%
This is achieved by partitioning the graph and sorting the edges by their
destination vertex. Thus, all updates that target the same vertex are processed
in the same batch. The size of the partitions is chosen such that the interval
of updates fits BRAM.

\subsubsection{Graph Compression}

Another way to reduce the amount of communication between the FPGA and the main
memory is graph compression~\cite{besta2018log, besta2018survey}. For example,
Lee et al.~\cite{lee2017extrav} use compression for accelerating graph algorithms. The
authors use their own compression scheme based on the established Webgraph
framework~\cite{boldi2004webgraph}.  They achieve compression ratios between
1.9 and 9.36, where the highest compression rates are obtained for web graphs.

\subsection{Insufficient FPGA Internal Memory}
\label{sec:problem_low-internal}

Modern FPGAs host a set of configurable memory modules, called BRAM, which
allow to store and access small amounts of data directly on the FPGA instead of
the DRAM on the host machine.  This memory can be compared to the cache in
modern CPUs, only that the FPGA decides which data is stored in BRAM.
Unfortunately, the capacity of BRAM is fairly small when compared to DRAM. 

The problem of insufficient amounts of BRAM to store full graph datasets is
complementary to the previous problem of insufficient bandwidth between FPGA
and DRAM:  fast and unlimited access to DRAM would diminish or even invalidate
the importance of the ability to store large data structures in BRAM (and vice
versa).  Thus, some of the approaches described in the following sections could
also be used to address the problem of insufficient DRAM--FPGA bandwidth.

Some existing approaches simply assume that  the whole graph data set can be
loaded into BRAM~\cite{babb1996solving, mencer2002hagar, jagadeesh2011field,
bondhugula:APSP_FPGA, kapre2006graphstep, weisz:GraphGen, engelhardt2016gravf}.
However, this approach limits the size of graphs that can be processed, see
Table~\ref{tab:papers} for the details on the largest graphs used in surveyed works.

\subsubsection{Partitioning the Graph}

A common solution for processing large data structures with insufficient storage space is 
to split the data structure into smaller partitions which are loaded and executed one after the 
other. Partitioning a graph is difficult because no partition is fully 
separated from others: While processing one partition, one might have to update vertices 
that are stored in other partitions that are still residing in the larger, slower memory.

Zhou and Prasanna use the edge-centric approach
of X-Stream but leave out the shuffle phase by directly storing the updates in
sets corresponding to the partitions~\cite{zhou2016high}. In another paper, the
same authors use the same model with some improvements, such as merging the
updates directly on the FPGA if the updates correspond to the same
partition~\cite{zhou2018framework}. In their approach, the graph is divided
into $k$ partitions of approximately similar sizes. Each partition consists of
an interval (range) of vertices, a shard containing edges that have one source
vertex in the interval, and a set of values representing the updates that come
from incoming edges (called \textbf{bins}). In the scatter phase, updates are
written to the corresponding bins. In the gather phase, these updates are
combined to compute new vertex values. 
Now, because of the data layout when processing one partition, all relevant
vertex properties can be loaded into BRAM.  Updates, however, can also target
vertices from other partitions and thus have to be written to DRAM during the
scatter phase. 

% In GraphGen~\cite{weisz:GraphGen, weisz:GraphGen_for_CoRAM}, the input graph is
% partitioned into groups with a similar number of vertices the set of edges
% that have their source vertex in that interval. Two partitions that have no
% connecting edges can be processed in parallel. 

Some works use more sophisticated partitioning approaches to improve the
locality of vertices inside each partition and thus reduce the inter-partition
communication. In GraphStep~\cite{kapre2006graphstep}, the authors use the
UMPack's multi-level partitioner~\cite{caldwell2000improved}. Another
work~\cite{kapre:custom_graph_FPGA} uses the hypergraph partitioning tool
PaToH~\cite{catalyurek1999hypergraph}. Elaborate partitioning can reduce
communication but it usually entails some additional pre-processing cost.

\subsubsection{Using multiple FPGAs}

% In one of the earlier works on graph FPGA computation in 2006, deLorimer et al.
% argue that a much higher performance can be achieved if the data resides in
% BRAM. They propose a system architecture called GraphStep, which consists of
% multiple connected FPGA devices. Each device features several processing units
% that are synchronized using a BSP like programming model. 

Many graph FPGA papers argue in favor of using multiple FPGAs and thus scaling
computational resources and achieving better performance~\cite{babb1996solving,
kapre2006graphstep, betkaoui2011framework, betkaoui2012reconfigurable,
attia2014cygraph, kapre:custom_graph_FPGA, dai:fpgp, dai:foregraph}. Some of
these papers require the use of multiple FPGA when processing larger graphs
that do not fit the BRAM modules of a single FPGA. Other designs aim to allow
the scaling of performance above the performance that can be delivered by a
single FPGA.
One common problem with using multiple FPGAs is the communication overhead.  In
FPGP~\cite{dai:fpgp}, the devices use a shared memory for the vertex properties
which poses a bandwidth problem if too many devices use the same memory.
ForeGraph~\cite{dai:foregraph} solves this problem by using a separate DRAM
module for each device and an interconnect between all devices through which
updates are transferred.

%       \begin{table}
%       \vspace{-1em}
%       \centering
%        \setlength{\tabcolsep}{2.5pt}
%       %\scriptsize
%       %\footnotesize
%       \ssmall
%       \sf
%       \begin{tabular}{@{}llllllll@{}}
%       %
%       \toprule
%       %
%       \textbf{Reference} & \textbf{Peak Bandwidth (GB/s)}   & \textbf{Random Read (ms)} & \textbf{Random Write (ms))} & \textbf{} & \textbf{Sequential Read(ms)} & \textbf{Scalable to} & \textbf{Max. Measured} \\
%       %
%       \midrule
%       %
%       DRAM              & 12-20            &                                       &                 &    & yes & n=220k, m=550k    &       \\
%       BRAM              & 5000-12000       &                                       &                 &   & yes & n=220k, m=550k    &       \\
%       HMC               & 60            &                                       &                 &    & yes & n=220k, m=550k    &       \\
%       HD                &              &                                       &                 &    & yes & n=220k, m=550k    &       \\
%       
%       % \cite{..}              &               &                                   &                       &       &     &                   &       \\
%       \bottomrule
%       %
%       \end{tabular}
%       %
%       \caption{Average memory access bandwidth and latencies for different memory technologies from an FPGA. }
%       %
%       \label{tab:access_times}
%       \vspace{-1em}
%       \end{table}

\section{Conclusion}

Graph processing on FPGAs is an important area of research as it can be used to accelerate numerous
graph algorithms by reducing the amount of consumed power. Yet, it contains a diverse set of
algorithms and processing frameworks, with a plethora of techniques and
approaches. We present the first survey that analyzes the rich world of graph processing on FPGAs.
We list and categorize the existing work, discuss key ideas, and present key insights and design choices.
Our work can be used by architects and developers
willing to select the best FPGA scheme in a given setting.

% \noindent
% \macb{ACKNOWLEDGEMENTS }
% %
% We thank ...

\renewcommand*{\bibfont}{\small}

% Bibliography
{\sf
%\scriptsize
%\bibliographystyle{ACM-Reference-Format}
\bibliographystyle{abbrv}
\bibliography{refs}

\begin{thebibliography}{100}

\bibitem{BRAM_power_consumption}
{{"Reducing Power Consumption in Xilinx FPGAs"}}.
\newblock available at: \url{https://vhdlguru.blogspot.com/2011/07}.

\bibitem{attia2014cygraph}
O.~G. Attia, T.~Johnson, K.~Townsend, P.~Jones, and J.~Zambreno.
\newblock Cygraph: A reconfigurable architecture for parallel breadth-first
  search.
\newblock In {\em 2014 IEEE International Parallel \& Distributed Processing
  Symposium Workshops (IPDPSW)}, pages 228--235. IEEE, 2014.

\bibitem{austin2010convey}
W.~Austin, V.~Heuveline, and J.-P. Weiss.
\newblock {\em Convey HC-1 hybrid core computer-The potential of FPGAs in
  numerical simulation}.
\newblock KIT, 2010.

\bibitem{babb1996solving}
J.~W. Babb, M.~Frank, and A.~Agarwal.
\newblock Solving graph problems with dynamic computation structures.
\newblock In {\em High-Speed Computing, Digital Signal Processing, and
  Filtering Using Reconfigurable Logic}, volume 2914, pages 225--237.
  International Society for Optics and Photonics, 1996.

\bibitem{chisel}
J.~Bachrach, H.~Vo, B.~Richards, Y.~Lee, A.~Waterman, R.~Avi{\v{z}}ienis,
  J.~Wawrzynek, and K.~Asanovi{\'c}.
\newblock Chisel: constructing hardware in a scala embedded language.
\newblock In {\em DAC Design Automation Conference 2012}, pages 1212--1221.
  IEEE, 2012.

\bibitem{beame1989optimal}
P.~Beame and J.~Hastad.
\newblock Optimal bounds for decision problems on the crcw pram.
\newblock {\em Journal of the ACM (JACM)}, 36(3):643--670, 1989.

\bibitem{beamer2013direction}
S.~Beamer, K.~Asanovi{\'c}, and D.~Patterson.
\newblock {{Direction-optimizing breadth-first search}}.
\newblock {\em Scientific Programming}, 21(3-4):137--148, 2013.

\bibitem{beamer2015gap}
S.~Beamer, K.~Asanovi{\'c}, and D.~Patterson.
\newblock The gap benchmark suite.
\newblock {\em arXiv preprint arXiv:1508.03619}, 2015.

\bibitem{bellman1958routing}
R.~Bellman.
\newblock On a routing problem.
\newblock {\em Quarterly of applied mathematics}, 16(1):87--90, 1958.

\bibitem{deep500}
T.~Ben-Nun, M.~Besta, S.~Huber, A.~N. Ziogas, D.~Peter, and T.~Hoefler.
\newblock {A Modular Benchmarking Infrastructure for High-Performance and
  Reproducible Deep Learning}.
\newblock IEEE, May 2019.
\newblock Accepted at the 33rd IEEE International Parallel \& Distributed
  Processing Symposium (IPDPS'19).

\bibitem{distdl-preprint}
T.~Ben-Nun and T.~Hoefler.
\newblock {Demystifying Parallel and Distributed Deep Learning: An In-Depth
  Concurrency Analysis}.
\newblock {\em CoRR}, abs/1802.09941, Feb. 2018.

\bibitem{bernaschi2012breadth}
M.~Bernaschi, M.~Bisson, E.~Mastrostefano, and D.~Rossetti.
\newblock Breadth first search on apenet+.
\newblock In {\em 2012 SC Companion: High Performance Computing, Networking
  Storage and Analysis}, pages 248--253. IEEE, 2012.

\bibitem{besta2019substream}
M.~Besta, M.~Fischer, T.~Ben-Nun, J.~D.~F. Licht, and T.~Hoefler.
\newblock {Substream-Centric Maximum Matchings on FPGA}.
\newblock Feb. 2019.
\newblock In Proceedings of the 27th ACM/SIGDA International Symposium on
  Field-Programmable Gate Arrays.

\bibitem{besta2018slim}
M.~Besta, S.~M. Hassan, S.~Yalamanchili, R.~Ausavarungnirun, O.~Mutlu, and
  T.~Hoefler.
\newblock Slim noc: A low-diameter on-chip network topology for high energy
  efficiency and scalability.
\newblock In {\em Proceedings of the Twenty-Third International Conference on
  Architectural Support for Programming Languages and Operating Systems}, pages
  43--55. ACM, 2018.

\bibitem{besta2014slim}
M.~Besta and T.~Hoefler.
\newblock Slim fly: A cost effective low-diameter network topology.
\newblock In {\em Proceedings of the International Conference for High
  Performance Computing, Networking, Storage and Analysis}, pages 348--359.
  IEEE Press, 2014.

\bibitem{besta2015accelerating}
M.~Besta and T.~Hoefler.
\newblock Accelerating irregular computations with hardware transactional
  memory and active messages.
\newblock In {\em Proceedings of the 24th International Symposium on
  High-Performance Parallel and Distributed Computing}, pages 161--172. ACM,
  2015.

\bibitem{besta2018survey}
M.~Besta and T.~Hoefler.
\newblock Survey and taxonomy of lossless graph compression and space-efficient
  graph representations.
\newblock {\em arXiv preprint arXiv:1806.01799}, 2018.

\bibitem{besta2017slimsell}
M.~Besta, F.~Marending, E.~Solomonik, and T.~Hoefler.
\newblock Slimsell: A vectorizable graph representation for breadth-first
  search.
\newblock In {\em Proc. IEEE IPDPS}, volume~17, 2017.

\bibitem{besta2017push}
M.~Besta, M.~Podstawski, L.~Groner, E.~Solomonik, and T.~Hoefler.
\newblock To push or to pull: On reducing communication and synchronization in
  graph computations.
\newblock In {\em Proceedings of the 26th International Symposium on
  High-Performance Parallel and Distributed Computing}, pages 93--104. ACM,
  2017.

\bibitem{besta2018log}
M.~Besta, D.~Stanojevic, T.~Zivic, J.~Singh, M.~Hoerold, and T.~Hoefler.
\newblock Log (graph): a near-optimal high-performance graph representation.
\newblock In {\em Proceedings of the 27th International Conference on Parallel
  Architectures and Compilation Techniques}, page~7. ACM, 2018.

\bibitem{betkaoui2011framework}
B.~Betkaoui, D.~B. Thomas, W.~Luk, and N.~Przulj.
\newblock A framework for fpga acceleration of large graph problems: Graphlet
  counting case study.
\newblock In {\em Field-Programmable Technology (FPT), 2011 International
  Conference on}, pages 1--8. IEEE, 2011.

\bibitem{betkaoui:APSP_FPGA}
B.~Betkaoui, Y.~Wang, D.~B. Thomas, and W.~Luk.
\newblock Parallel fpga-based all pairs shortest paths for sparse networks: A
  human brain connectome case study.
\newblock In {\em 22nd International Conference on Field Programmable Logic and
  Applications (FPL)}, pages 99--104, Aug 2012.

\bibitem{betkaoui2012reconfigurable}
B.~Betkaoui, Y.~Wang, D.~B. Thomas, and W.~Luk.
\newblock A reconfigurable computing approach for efficient and scalable
  parallel graph exploration.
\newblock In {\em Application-Specific Systems, Architectures and Processors
  (ASAP), 2012 IEEE 23rd International Conference on}, pages 8--15. IEEE, 2012.

\bibitem{boldi2004webgraph}
P.~Boldi and S.~Vigna.
\newblock The webgraph framework i: compression techniques.
\newblock In {\em Proceedings of the 13th international conference on World
  Wide Web}, pages 595--602. ACM, 2004.

\bibitem{boman2013scalable}
E.~G. Boman, K.~D. Devine, and S.~Rajamanickam.
\newblock Scalable matrix computations on large scale-free graphs using 2d
  graph partitioning.
\newblock In {\em SC'13: Proceedings of the International Conference on High
  Performance Computing, Networking, Storage and Analysis}, pages 1--12. IEEE,
  2013.

\bibitem{bondhugula2006hardware}
U.~Bondhugula, A.~Devulapalli, J.~Dinan, J.~Fernando, P.~Wyckoff, E.~Stahlberg,
  and P.~Sadayappan.
\newblock Hardware/software integration for fpga-based all-pairs
  shortest-paths.
\newblock In {\em Field-Programmable Custom Computing Machines, 2006. FCCM'06.
  14th Annual IEEE Symposium on}, pages 152--164. IEEE, 2006.

\bibitem{bondhugula:APSP_FPGA}
U.~Bondhugula, A.~Devulapalli, J.~Fernando, P.~Wyckoff, and P.~Sadayappan.
\newblock Parallel fpga-based all-pairs shortest-paths in a directed graph.
\newblock In {\em Proceedings 20th IEEE International Parallel Distributed
  Processing Symposium}, pages 10 pp.--, April 2006.

\bibitem{boruuvka1926jistem}
O.~Boruvka.
\newblock {O jist{\'e}m probl{\'e}mu minim{\'a}ln{\'\i}m}.
\newblock 1926.

\bibitem{brandes2001faster}
U.~Brandes.
\newblock {A faster algorithm for betweenness centrality*}.
\newblock {\em Journal of Mathematical Sociology}, 25(2):163--177, 2001.

\bibitem{caceres1997efficient}
E.~C{\'a}ceres, F.~Dehne, A.~Ferreira, P.~Flocchini, I.~Rieping, A.~Roncato,
  N.~Santoro, and S.~W. Song.
\newblock Efficient parallel graph algorithms for coarse grained multicomputers
  and bsp.
\newblock In {\em International Colloquium on Automata, Languages, and
  Programming}, pages 390--400. Springer, 1997.

\bibitem{caldwell2000improved}
A.~E. Caldwell, A.~B. Kahng, and I.~L. Markov.
\newblock Improved algorithms for hypergraph bipartitioning.
\newblock In {\em Proceedings of the 2000 Asia and South Pacific Design
  Automation Conference}, pages 661--666. ACM, 2000.

\bibitem{catalyurek1999hypergraph}
U.~V. Catalyurek and C.~Aykanat.
\newblock Hypergraph-partitioning based decomposition for parallel
  sparse-matrix vector multiplication.
\newblock {\em IEEE Transactions on parallel and distributed systems},
  10(7):673--693, 1999.

\bibitem{chazelle2000minimum}
B.~Chazelle.
\newblock A minimum spanning tree algorithm with inverse-ackermann type
  complexity.
\newblock {\em Journal of the ACM (JACM)}, 47(6):1028--1047, 2000.

\bibitem{chi2016nxgraph}
Y.~Chi, G.~Dai, Y.~Wang, G.~Sun, G.~Li, and H.~Yang.
\newblock Nxgraph: An efficient graph processing system on a single machine.
\newblock In {\em Data Engineering (ICDE), 2016 IEEE 32nd International
  Conference on}, pages 409--420. IEEE, 2016.

\bibitem{ching2015one}
A.~Ching, S.~Edunov, M.~Kabiljo, D.~Logothetis, and S.~Muthukrishnan.
\newblock One trillion edges: Graph processing at facebook-scale.
\newblock {\em Proceedings of the VLDB Endowment}, 8(12):1804--1815, 2015.

\bibitem{chung2011coram}
E.~S. Chung, J.~C. Hoe, and K.~Mai.
\newblock Coram: an in-fabric memory architecture for fpga-based computing.
\newblock In {\em Proceedings of the 19th ACM/SIGDA international symposium on
  Field programmable gate arrays}, pages 97--106. ACM, 2011.

\bibitem{cohen2009graph}
J.~Cohen.
\newblock Graph twiddling in a mapreduce world.
\newblock {\em Computing in Science \& Engineering}, 11(4):29--41, 2009.

\bibitem{compeau2011apply}
P.~E. Compeau, P.~A. Pevzner, and G.~Tesler.
\newblock How to apply de bruijn graphs to genome assembly.
\newblock {\em Nature biotechnology}, 29(11):987, 2011.

\bibitem{Cormen:2001:IA:580470}
T.~H. Cormen, C.~Stein, R.~L. Rivest, and C.~E. Leiserson.
\newblock {\em {Introduction to Algorithms}}.
\newblock McGraw-Hill Higher Education, 2nd edition, 2001.

\bibitem{dai:fpgp}
G.~Dai, Y.~Chi, Y.~Wang, and H.~Yang.
\newblock Fpgp: Graph processing framework on fpga a case study of
  breadth-first search.
\newblock In {\em Proceedings of the 2016 ACM/SIGDA International Symposium on
  Field-Programmable Gate Arrays}, FPGA '16, pages 105--110, New York, NY, USA,
  2016. ACM.

\bibitem{dai:foregraph}
G.~Dai, T.~Huang, Y.~Chi, N.~Xu, Y.~Wang, and H.~Yang.
\newblock Foregraph: Exploring large-scale graph processing on multi-fpga
  architecture.
\newblock In {\em Proceedings of the 2017 ACM/SIGDA International Symposium on
  Field-Programmable Gate Arrays}, FPGA '17, pages 217--226, New York, NY, USA,
  2017. ACM.

\bibitem{dally2004principles}
W.~J. Dally and B.~P. Towles.
\newblock {\em Principles and practices of interconnection networks}.
\newblock Elsevier, 2004.

\bibitem{dandalis1999domain}
A.~Dandalis, A.~Mei, and V.~K. Prasanna.
\newblock Domain specific mapping for solving graph problems on reconfigurable
  devices.
\newblock In {\em International Parallel Processing Symposium}, pages 652--660.
  Springer, 1999.

\bibitem{dean2008mapreduce}
J.~Dean and S.~Ghemawat.
\newblock {MapReduce: simplified data processing on large clusters}.
\newblock {\em Communications of the ACM}, 51(1):107--113, 2008.

\bibitem{dijkstra1959note}
E.~W. Dijkstra.
\newblock A note on two problems in connexion with graphs.
\newblock {\em Numerische mathematik}, 1(1):269--271, 1959.

\bibitem{ediger2013investigating}
D.~Ediger and D.~A. Bader.
\newblock Investigating graph algorithms in the bsp model on the cray xmt.
\newblock In {\em Parallel and Distributed Processing Symposium Workshops \&
  PhD Forum (IPDPSW), 2013 IEEE 27th International}, pages 1638--1645. IEEE,
  2013.

\bibitem{edmonds1965paths}
J.~Edmonds.
\newblock Paths, trees, and flowers.
\newblock {\em Canadian Journal of mathematics}, 17(3):449--467, 1965.

\bibitem{edmonds2006single}
N.~Edmonds, A.~Breuer, D.~P. Gregor, and A.~Lumsdaine.
\newblock Single-source shortest paths with the parallel boost graph library.
\newblock In {\em The Shortest Path Problem}, pages 219--248, 2006.

\bibitem{engelhardt2016gravf}
N.~Engelhardt and H.~K.-H. So.
\newblock Gravf: A vertex-centric distributed graph processing framework on
  fpgas.
\newblock In {\em Field Programmable Logic and Applications (FPL), 2016 26th
  International Conference on}, pages 1--4. IEEE, 2016.

\bibitem{floyd1962algorithm}
R.~W. Floyd.
\newblock Algorithm 97: shortest path.
\newblock {\em Communications of the ACM}, 5(6):345, 1962.

\bibitem{ford1956network}
L.~R. Ford~Jr.
\newblock Network flow theory.
\newblock Technical report, RAND CORP SANTA MONICA CA, 1956.

\bibitem{fredman1987fibonacci}
M.~L. Fredman and R.~E. Tarjan.
\newblock Fibonacci heaps and their uses in improved network optimization
  algorithms.
\newblock {\em Journal of the ACM (JACM)}, 34(3):596--615, 1987.

\bibitem{gao2016hrl}
M.~Gao and C.~Kozyrakis.
\newblock Hrl: Efficient and flexible reconfigurable logic for near-data
  processing.
\newblock In {\em High Performance Computer Architecture (HPCA), 2016 IEEE
  International Symposium on}, pages 126--137. Ieee, 2016.

\bibitem{gerstenberger2014enabling}
R.~Gerstenberger, M.~Besta, and T.~Hoefler.
\newblock Enabling highly-scalable remote memory access programming with mpi-3
  one sided.
\newblock {\em Scientific Programming}, 22(2):75--91, 2014.

\bibitem{gianinazzi2018communication}
L.~Gianinazzi, P.~Kalvoda, A.~De~Palma, M.~Besta, and T.~Hoefler.
\newblock Communication-avoiding parallel minimum cuts and connected
  components.
\newblock In {\em Proceedings of the 23rd ACM SIGPLAN Symposium on Principles
  and Practice of Parallel Programming}, pages 219--232. ACM, 2018.

\bibitem{gibbons1989more}
P.~B. Gibbons.
\newblock A more practical pram model.
\newblock In {\em Proceedings of the first annual ACM symposium on Parallel
  algorithms and architectures}, pages 158--168. ACM, 1989.

\bibitem{Han1997}
Y.~Han, V.~Y. Pan, and J.~H. Reif.
\newblock Efficient parallel algorithms for computing all pair shortest paths
  in directed graphs.
\newblock {\em Algorithmica}, 17(4):399--415, Apr 1997.

\bibitem{hauck2010reconfigurable}
S.~Hauck and A.~DeHon.
\newblock {\em Reconfigurable computing: the theory and practice of FPGA-based
  computation}, volume~1.
\newblock Elsevier, 2010.

\bibitem{herlihy1993transactional}
M.~Herlihy and J.~E.~B. Moss.
\newblock {\em Transactional memory: Architectural support for lock-free data
  structures}, volume~21.
\newblock ACM, 1993.

\bibitem{hong2014simplifying}
S.~Hong, S.~Salihoglu, J.~Widom, and K.~Olukotun.
\newblock Simplifying scalable graph processing with a domain-specific
  language.
\newblock In {\em Proceedings of Annual IEEE/ACM International Symposium on
  Code Generation and Optimization}, page 208. ACM, 2014.

\bibitem{horowitz_energy}
M.~Horowitz.
\newblock 1.1 computing's energy problem (and what we can do about it).
\newblock In {\em 2014 IEEE International Solid-State Circuits Conference
  Digest of Technical Papers (ISSCC)}, pages 10--14, Feb 2014.

\bibitem{huelsbergen2000representation}
L.~Huelsbergen.
\newblock A representation for dynamic graphs in reconfigurable hardware and
  its application to fundamental graph algorithms.
\newblock In {\em Proceedings of the 2000 ACM/SIGDA eighth international
  symposium on Field programmable gate arrays}, pages 105--115. ACM, 2000.

\bibitem{hyperflex}
Intel.
\newblock {Understanding How the New Intel HyperFlex {FPGA} Architecture
  Enables Next-Generation High-Performance Systems}.
\newblock Technical report.

\bibitem{jacob2010memory}
B.~Jacob, S.~Ng, and D.~Wang.
\newblock {\em Memory systems: cache, DRAM, disk}.
\newblock Morgan Kaufmann, 2010.

\bibitem{jagadeesh2011field}
G.~R. Jagadeesh, T.~Srikanthan, and C.~Lim.
\newblock Field programmable gate array-based acceleration of shortest-path
  computation.
\newblock {\em IET computers \& digital techniques}, 5(4):231--237, 2011.

\bibitem{jiang2011short}
B.~Jiang.
\newblock A short note on data-intensive geospatial computing.
\newblock In {\em Information Fusion and Geographic Information Systems}, pages
  13--17. Springer, 2011.

\bibitem{johnson1977efficient}
D.~B. Johnson.
\newblock {Efficient algorithms for shortest paths in sparse networks}.
\newblock {\em Journal of the ACM (JACM)}, 24(1):1--13, 1977.

\bibitem{kalavri2018high}
V.~Kalavri, V.~Vlassov, and S.~Haridi.
\newblock High-level programming abstractions for distributed graph processing.
\newblock {\em IEEE Transactions on Knowledge and Data Engineering},
  30(2):305--324, 2018.

\bibitem{kaleem15sgd}
R.~Kaleem, S.~Pai, and K.~Pingali.
\newblock Stochastic gradient descent on gpus.
\newblock In {\em Proceedings of the 8th Workshop on General Purpose Processing
  Using GPUs}, GPGPU-8, pages 81--89, New York, NY, USA, 2015. ACM.

\bibitem{kapre:custom_graph_FPGA}
N.~Kapre.
\newblock Custom fpga-based soft-processors for sparse graph acceleration.
\newblock In {\em 2015 IEEE 26th International Conference on
  Application-specific Systems, Architectures and Processors (ASAP)}, pages
  9--16, July 2015.

\bibitem{kapre2006graphstep}
N.~Kapre, N.~Mehta, D.~Rizzo, I.~Eslick, R.~Rubin, T.~E. Uribe, F.~Thomas~Jr,
  A.~DeHon, et~al.
\newblock Graphstep: A system architecture for sparse-graph algorithms.
\newblock In {\em Field-Programmable Custom Computing Machines, 2006. FCCM'06.
  14th Annual IEEE Symposium on}, pages 143--151. IEEE, 2006.

\bibitem{kapre:SoCs_FPGA}
N.~Kapre and P.~Moorthy.
\newblock A case for embedded fpga-based socs in energy-efficient acceleration
  of graph problems.
\newblock {\em Supercomput. Front. Innov.: Int. J.}, 2(3):76--86, July 2015.

\bibitem{kepner2016mathematical}
J.~Kepner, P.~Aaltonen, D.~Bader, A.~Bulu{\c{c}}, F.~Franchetti, J.~Gilbert,
  D.~Hutchison, M.~Kumar, A.~Lumsdaine, H.~Meyerhenke, et~al.
\newblock Mathematical foundations of the graphblas.
\newblock {\em arXiv preprint arXiv:1606.05790}, 2016.

\bibitem{kepner2011graph}
J.~Kepner and J.~Gilbert.
\newblock {\em {Graph algorithms in the language of linear algebra}},
  volume~22.
\newblock SIAM, 2011.

\bibitem{khan2016query}
A.~Khan and C.~Aggarwal.
\newblock {Query-friendly compression of graph streams}.
\newblock In {\em Advances in Social Networks Analysis and Mining (ASONAM),
  2016 IEEE/ACM International Conference on}, pages 130--137. IEEE, 2016.

\bibitem{khoram2018accelerating}
S.~Khoram, J.~Zhang, M.~Strange, and J.~Li.
\newblock Accelerating graph analytics by co-optimizing storage and access on
  an fpga-hmc platform.
\newblock In {\em Proceedings of the 2018 ACM/SIGDA International Symposium on
  Field-Programmable Gate Arrays}, pages 239--248. ACM, 2018.

\bibitem{kruskal1956shortest}
J.~B. Kruskal.
\newblock On the shortest spanning subtree of a graph and the traveling
  salesman problem.
\newblock {\em Proceedings of the American Mathematical society}, 7(1):48--50,
  1956.

\bibitem{kyrola2012graphchi}
A.~Kyrola, G.~Blelloch, and C.~Guestrin.
\newblock {GraphChi: large-scale graph computation on just a PC}.
\newblock In {\em Presented as part of the 10th USENIX Symposium on Operating
  Systems Design and Implementation (OSDI 12)}, pages 31--46, 2012.

\bibitem{lee2017extrav}
J.~Lee, H.~Kim, S.~Yoo, K.~Choi, H.~P. Hofstee, G.-J. Nam, M.~R. Nutter, and
  D.~Jamsek.
\newblock Extrav: boosting graph processing near storage with a coherent
  accelerator.
\newblock {\em Proceedings of the VLDB Endowment}, 10(12):1706--1717, 2017.

\bibitem{lei2016fpga}
G.~Lei, Y.~Dou, R.~Li, and F.~Xia.
\newblock An fpga implementation for solving the large
  single-source-shortest-path problem.
\newblock {\em IEEE Transactions on Circuits and Systems II: Express Briefs},
  63(5):473--477, 2016.

\bibitem{leiserson2010work}
C.~E. Leiserson and T.~B. Schardl.
\newblock A work-efficient parallel breadth-first search algorithm (or how to
  cope with the nondeterminism of reducers).
\newblock In {\em Proceedings of the twenty-second annual ACM symposium on
  Parallelism in algorithms and architectures}, pages 303--314. ACM, 2010.

\bibitem{liu2004conceptnet}
H.~Liu and P.~Singh.
\newblock Conceptnet—a practical commonsense reasoning tool-kit.
\newblock {\em BT technology journal}, 22(4):211--226, 2004.

\bibitem{low2010graphlab}
Y.~Low, J.~Gonzalez, A.~Kyrola, D.~Bickson, C.~Guestrin, and J.~M. Hellerstein.
\newblock {{Graphlab: A new framework for parallel machine learning}}.
\newblock {\em preprint arXiv:1006.4990}, 2010.

\bibitem{DBLP:journals/ppl/LumsdaineGHB07}
A.~Lumsdaine, D.~Gregor, B.~Hendrickson, and J.~W. Berry.
\newblock {Challenges in Parallel Graph Processing}.
\newblock {\em Par. Proc. Let.}, 17(1):5--20, 2007.

\bibitem{ma2017fpga}
X.~Ma, D.~Zhang, and D.~Chiou.
\newblock Fpga-accelerated transactional execution of graph workloads.
\newblock In {\em Proceedings of the 2017 ACM/SIGDA International Symposium on
  Field-Programmable Gate Arrays}, pages 227--236. ACM, 2017.

\bibitem{malewicz2010pregel}
G.~Malewicz, M.~H. Austern, A.~J. Bik, J.~C. Dehnert, I.~Horn, N.~Leiser, and
  G.~Czajkowski.
\newblock Pregel: a system for large-scale graph processing.
\newblock In {\em Proceedings of the 2010 ACM SIGMOD International Conference
  on Management of data}, pages 135--146. ACM, 2010.

\bibitem{mencer2002hagar}
O.~Mencer, Z.~Huang, and L.~Huelsbergen.
\newblock Hagar: Efficient multi-context graph processors.
\newblock In {\em International Conference on Field Programmable Logic and
  Applications}, pages 915--924. Springer, 2002.

\bibitem{meyer2003delta}
U.~Meyer and P.~Sanders.
\newblock {{$\Delta$-stepping: a parallelizable shortest path algorithm}}.
\newblock {\em Journal of Algorithms}, 49(1):114--152, 2003.

\bibitem{hls_survey}
R.~Nane, V.~Sima, C.~Pilato, J.~Choi, B.~Fort, A.~Canis, Y.~T. Chen, H.~Hsiao,
  S.~Brown, F.~Ferrandi, J.~Anderson, and K.~Bertels.
\newblock A survey and evaluation of fpga high-level synthesis tools.
\newblock {\em IEEE Transactions on Computer-Aided Design of Integrated
  Circuits and Systems}, 35(10):1591--1604, Oct 2016.

\bibitem{newman2005measure}
M.~E. Newman.
\newblock A measure of betweenness centrality based on random walks.
\newblock {\em Social networks}, 27(1):39--54, 2005.

\bibitem{ni2014parallel}
S.~Ni, Y.~Dou, D.~Zou, R.~Li, and Q.~Wang.
\newblock Parallel graph traversal for fpga.
\newblock {\em IEICE Electronics Express}, 11(7):20130987--20130987, 2014.

\bibitem{weisz:GraphGen}
E.~Nurvitadhi, G.~Weisz, Y.~Wang, S.~Hurkat, M.~Nguyen, J.~C. Hoe, J.~F.
  Martínez, and C.~Guestrin.
\newblock Graphgen: An fpga framework for vertex-centric graph computation.
\newblock In {\em 2014 IEEE 22nd Annual International Symposium on
  Field-Programmable Custom Computing Machines}, pages 25--28, May 2014.

\bibitem{oguntebi:GraphOps}
T.~Oguntebi and K.~Olukotun.
\newblock Graphops: A dataflow library for graph analytics acceleration.
\newblock In {\em Proceedings of the 2016 ACM/SIGDA International Symposium on
  Field-Programmable Gate Arrays}, FPGA '16, pages 111--117, New York, NY, USA,
  2016. ACM.

\bibitem{ozdal2016energy}
M.~M. Ozdal, S.~Yesil, T.~Kim, A.~Ayupov, J.~Greth, S.~Burns, and O.~Ozturk.
\newblock Energy efficient architecture for graph analytics accelerators.
\newblock In {\em Computer Architecture (ISCA), 2016 ACM/IEEE 43rd Annual
  International Symposium on}, pages 166--177. IEEE, 2016.

\bibitem{o2016neural}
M.~O’Neill.
\newblock Neural network for recognition of handwritten digits (2006).
\newblock {\em Source: http://www. codeproject.
  com/KB/library/NeuralNetRecognition. asp x}, 2016.

\bibitem{page1999pagerank}
L.~Page, S.~Brin, R.~Motwani, and T.~Winograd.
\newblock The pagerank citation ranking: Bringing order to the web.
\newblock Technical report, Stanford InfoLab, 1999.

\bibitem{papadimitriou1998combinatorial}
C.~H. Papadimitriou and K.~Steiglitz.
\newblock {\em Combinatorial optimization: algorithms and complexity}.
\newblock Courier Corporation, 1998.

\bibitem{pawlowski2011hybrid}
J.~T. Pawlowski.
\newblock Hybrid memory cube (hmc).
\newblock In {\em 2011 IEEE Hot chips 23 symposium (HCS)}, pages 1--24. IEEE,
  2011.

\bibitem{maxeler}
O.~Pell, O.~Mencer, K.~H. Tsoi, and W.~Luk.
\newblock Maximum performance computing with dataflow engines.
\newblock In {\em High-performance computing using FPGAs}, pages 747--774.
  Springer, 2013.

\bibitem{poirier2018dna}
C.~Poirier, B.~Gosselin, and P.~Fortier.
\newblock Dna assembly with de bruijn graphs using an fpga platform.
\newblock {\em IEEE/ACM transactions on computational biology and
  bioinformatics}, 15(3):1003--1009, 2018.

\bibitem{prim1957shortest}
R.~C. Prim.
\newblock Shortest connection networks and some generalizations.
\newblock {\em Bell system technical journal}, 36(6):1389--1401, 1957.

\bibitem{roy2013x}
A.~Roy, I.~Mihailovic, and W.~Zwaenepoel.
\newblock X-stream: Edge-centric graph processing using streaming partitions.
\newblock In {\em Proceedings of the Twenty-Fourth ACM Symposium on Operating
  Systems Principles}, pages 472--488. ACM, 2013.

\bibitem{saad1990sparskit}
Y.~Saad.
\newblock Sparskit: A basic tool kit for sparse matrix computations.
\newblock 1990.

\bibitem{schank2007algorithmic}
T.~Schank.
\newblock Algorithmic aspects of triangle-based network analysis.
\newblock 2007.

\bibitem{schmid2016high}
P.~Schmid, M.~Besta, and T.~Hoefler.
\newblock High-performance distributed rma locks.
\newblock In {\em Proceedings of the 25th ACM International Symposium on
  High-Performance Parallel and Distributed Computing}, pages 19--30. ACM,
  2016.

\bibitem{schweizer2015evaluating}
H.~Schweizer, M.~Besta, and T.~Hoefler.
\newblock Evaluating the cost of atomic operations on modern architectures.
\newblock In {\em 2015 International Conference on Parallel Architecture and
  Compilation (PACT)}, pages 445--456. IEEE, 2015.

\bibitem{shiloach1980log}
Y.~Shiloach and U.~Vishkin.
\newblock An o (log n) parallel connectivity algorithm.
\newblock Technical report, Computer Science Department, Technion, 1980.

\bibitem{shun2015multicore}
J.~Shun and K.~Tangwongsan.
\newblock Multicore triangle computations without tuning.
\newblock In {\em Data Engineering (ICDE), 2015 IEEE 31st International
  Conference on}, pages 149--160. IEEE, 2015.

\bibitem{solomonik2017scaling}
E.~Solomonik, M.~Besta, F.~Vella, and T.~Hoefler.
\newblock Scaling betweenness centrality using communication-efficient sparse
  matrix multiplication.
\newblock In {\em Proceedings of the International Conference for High
  Performance Computing, Networking, Storage and Analysis}, page~47. ACM, 2017.

\bibitem{sridharan2009hardware}
K.~Sridharan, T.~Priya, and P.~R. Kumar.
\newblock Hardware architecture for finding shortest paths.
\newblock In {\em TENCON 2009-2009 IEEE Region 10 Conference}, pages 1--5.
  IEEE, 2009.

\bibitem{sun2006accelerating}
C.~Sun, E.~W. Chew, N.~Shaikh~Husin, and M.~Khalil-Hani.
\newblock Accelerating graph algorithms with priority queue processor.
\newblock In {\em Regional Postgraduate Conference on Engineering and Science
  (RPCES 2006)}, pages 257--262, 2006.

\bibitem{sun2003stereo}
J.~Sun, N.-N. Zheng, and H.-Y. Shum.
\newblock Stereo matching using belief propagation.
\newblock {\em IEEE Transactions on pattern analysis and machine intelligence},
  25(7):787--800, 2003.

\bibitem{szeliski2008comparative}
R.~Szeliski, R.~Zabih, D.~Scharstein, O.~Veksler, V.~Kolmogorov, A.~Agarwala,
  M.~Tappen, and C.~Rother.
\newblock A comparative study of energy minimization methods for markov random
  fields with smoothness-based priors.
\newblock {\em IEEE transactions on pattern analysis and machine intelligence},
  30(6):1068--1080, 2008.

\bibitem{tate2014programming}
A.~Tate, A.~Kamil, A.~Dubey, A.~Gr{\"o}{\ss}linger, B.~Chamberlain, B.~Goglin,
  C.~Edwards, C.~J. Newburn, D.~Padua, D.~Unat, et~al.
\newblock Programming abstractions for data locality.
\newblock PADAL Workshop 2014, April 28--29, Swiss National Supercomputing
  Center~…, 2014.

\bibitem{thorup1999undirected}
M.~Thorup.
\newblock Undirected single-source shortest paths with positive integer weights
  in linear time.
\newblock {\em Journal of the ACM (JACM)}, 46(3):362--394, 1999.

\bibitem{tommiska2001dijkstra}
M.~Tommiska and J.~Skytt{\"a}.
\newblock Dijkstra’s shortest path routing algorithm in reconfigurable
  hardware.
\newblock In {\em International Conference on Field Programmable Logic and
  Applications}, pages 653--657. Springer, 2001.

\bibitem{umuroglu:hybrid_bfs_FPGA}
Y.~Umuroglu, D.~Morrison, and M.~Jahre.
\newblock Hybrid breadth-first search on a single-chip fpga-cpu heterogeneous
  platform.
\newblock In {\em 2015 25th International Conference on Field Programmable
  Logic and Applications (FPL)}, pages 1--8, Sept 2015.

\bibitem{valiant1990bridging}
L.~G. Valiant.
\newblock {A bridging model for parallel computation}.
\newblock {\em Communications of the ACM}, 33(8):103--111, 1990.

\bibitem{varma2013fassem}
B.~S.~C. Varma, K.~Paul, M.~Balakrishnan, and D.~Lavenier.
\newblock Fassem: Fpga based acceleration of de novo genome assembly.
\newblock In {\em Field-Programmable Custom Computing Machines (FCCM), 2013
  IEEE 21st Annual International Symposium on}, pages 173--176. IEEE, 2013.

\bibitem{venkataraman2003blocked}
G.~Venkataraman, S.~Sahni, and S.~Mukhopadhyaya.
\newblock A blocked all-pairs shortest-paths algorithm.
\newblock {\em Journal of Experimental Algorithmics (JEA)}, 8:2--2, 2003.

\bibitem{wang2010message}
Q.~Wang, W.~Jiang, Y.~Xia, and V.~Prasanna.
\newblock A message-passing multi-softcore architecture on fpga for
  breadth-first search.
\newblock In {\em Field-Programmable Technology (FPT), 2010 International
  Conference on}, pages 70--77. IEEE, 2010.

\bibitem{watts1998collective}
D.~J. Watts and S.~H. Strogatz.
\newblock Collective dynamics of ‘small-world’networks.
\newblock {\em nature}, 393(6684):440, 1998.

\bibitem{ultraram}
Xilinx.
\newblock {UltraRAM: Breakthrough Embedded Memory Integration on UltraScale+
  Devices}.
\newblock Technical report, 06 2016.

\bibitem{versal}
Xilinx.
\newblock {Versal: The First Adaptive Compute Acceleration Platform (ACAP)}.
\newblock Technical report, 02 2018.

\bibitem{yang2018efficient}
C.~Yang.
\newblock An efficient dispatcher for large scale graphprocessing on
  opencl-based fpgas.
\newblock {\em arXiv preprint arXiv:1806.11509}, 2018.

\bibitem{yang2015fast}
C.~Yang, Y.~Wang, and J.~D. Owens.
\newblock {Fast sparse matrix and sparse vector multiplication algorithm on the
  GPU}.
\newblock In {\em Par. and Dist. Proc. Symp. Work. (IPDPSW), IEEE Intl.}, pages
  841--847. IEEE, 2015.

\bibitem{yao2018efficient}
P.~Yao.
\newblock An efficient graph accelerator with parallel data conflict
  management.
\newblock {\em arXiv preprint arXiv:1806.00751}, 2018.

\bibitem{horawalavithana:graph}
Y.S.Horawalavithana.
\newblock {On the Design of an Efficient Hardware Accelerator for Large Scale
  Graph Analytics}.

\bibitem{zhang:graph_FPGA}
J.~Zhang, S.~Khoram, and J.~Li.
\newblock Boosting the performance of fpga-based graph processor using hybrid
  memory cube: A case for breadth first search.
\newblock In {\em Proceedings of the 2017 ACM/SIGDA International Symposium on
  Field-Programmable Gate Arrays}, FPGA '17, pages 207--216, New York, NY, USA,
  2017. ACM.

\bibitem{zhang2018degree}
J.~Zhang and J.~Li.
\newblock Degree-aware hybrid graph traversal on fpga-hmc platform.
\newblock In {\em Proceedings of the 2018 ACM/SIGDA International Symposium on
  Field-Programmable Gate Arrays}, pages 229--238. ACM, 2018.

\bibitem{cambriconx}
S.~{Zhang}, Z.~{Du}, L.~{Zhang}, H.~{Lan}, S.~{Liu}, L.~{Li}, Q.~{Guo},
  T.~{Chen}, and Y.~{Chen}.
\newblock Cambricon-x: An accelerator for sparse neural networks.
\newblock In {\em 2016 49th Annual IEEE/ACM International Symposium on
  Microarchitecture (MICRO)}, pages 1--12, Oct 2016.

\bibitem{zhou2015sssp}
S.~Zhou, C.~Chelmis, and V.~K. Prasanna.
\newblock Accelerating large-scale single-source shortest path on fpga.
\newblock In {\em Parallel and Distributed Processing Symposium Workshop
  (IPDPSW), 2015 IEEE International}, pages 129--136. IEEE, 2015.

\bibitem{zhou2015pagerank}
S.~Zhou, C.~Chelmis, and V.~K. Prasanna.
\newblock Optimizing memory performance for fpga implementation of pagerank.
\newblock In {\em ReConFig}, pages 1--6, 2015.

\bibitem{zhou2016high}
S.~Zhou, C.~Chelmis, and V.~K. Prasanna.
\newblock High-throughput and energy-efficient graph processing on fpga.
\newblock In {\em Field-Programmable Custom Computing Machines (FCCM), 2016
  IEEE 24th Annual International Symposium on}, pages 103--110. IEEE, 2016.

\bibitem{zhou2018framework}
S.~Zhou, R.~Kannan, H.~Zeng, and V.~K. Prasanna.
\newblock An fpga framework for edge-centric graph processing.
\newblock In {\em Proceedings of the 15th ACM International Conference on
  Computing Frontiers}, pages 69--77. ACM, 2018.

\bibitem{zhou2017accelerating}
S.~Zhou and V.~K. Prasanna.
\newblock Accelerating graph analytics on cpu-fpga heterogeneous platform.
\newblock In {\em 2017 29th International Symposium on Computer Architecture
  and High Performance Computing (SBAC-PAD)}, pages 137--144. IEEE, 2017.

\end{thebibliography}
%\bibliography{../refs_ultrashort}
}

\end{document}